\documentclass[twocolumn,showpacs,amsmath,amssymb,prb]{revtex4}
\usepackage{dcolumn}
\usepackage{bm}
\usepackage[dvips]{graphicx}
\def\psym#1{}
\def\abs#1{\left\vert#1\right\vert}

\begin{document}

\title{Exchange interactions and high-energy spin states in 
$\rm \bf Mn_{12}$-acetate}

\author{G. Chaboussant}
\altaffiliation{Present address: Laboratoire L\'eon Brillouin, CEA 
Saclay, 91191 Gif-sur-Yvette Cedex, France}
\author{A. Sieber}
\author{S. Ochsenbein}
\author{H.-U. G\"udel}
\author{M. Murrie}
\affiliation{Departement f\"ur Chemie und Biochemie, Universit\"at Bern,
Freiestrasse 3, CH-3000 Bern 9, Switzerland}

\author{A. Honecker}
\affiliation{Institut f\"ur Theoretische Physik, Technische Universit\"at 
Braunschweig, Mendelssohnstr.~3, 38106 Braunschweig, Germany}

\author{N. Fukushima}
\affiliation{Institut f\"ur Theoretische Physik, Technische Universit\"at 
Braunschweig, Mendelssohnstr.~3, 38106 Braunschweig, Germany}

\author{B. Normand}
\altaffiliation{Corresponding author: bruce.normand@unifr.ch}
\affiliation{D\'epartement de Physique, Universit\'e de Fribourg, 
CH-1700 Fribourg, Switzerland}

\date{\today}

\begin{abstract}

We perform inelastic neutron scattering measurements on the molecular 
nanomagnet Mn$_{12}$-acetate to measure the excitation spectrum up to 
45 meV (500 K). We isolate magnetic excitations in two groups at 5--6.5 
meV (60--75 K) and 8--10.5 meV (95--120 K), with higher levels appearing 
only at 27 meV (310 K) and 31 meV (360 K). From a detailed 
characterization of the transition peaks we show that all of the 
low-energy modes appear to be separate $S$ = 9 excitations above the 
$S$ = 10 ground state, with the peak at 27 meV (310 K) corresponding to 
the first $S$ = 11 excitation. We consider a general model for the four 
exchange interaction parameters of the molecule. The static susceptibility 
is computed by high-temperature series expansion and the energy spectrum, 
matrix elements and ground-state spin configuration by exact diagonalization. 
The theoretical results are matched with experimental observation by 
inclusion of cluster anisotropy parameters, revealing strong constraints 
on possible parameter sets. We conclude that only a model with dominant 
exchange couplings $J_1 \sim J_2 \sim 5.5$ meV (65 K) and small couplings 
$J_3 \sim J_4 \sim 0.6$ meV (7 K) is consistent with the experimental data. 

\end{abstract}

\pacs{75.30.Et, 75.50.Xx, 78.70.Nx}

\maketitle

\section{Introduction}

The topic of molecular magnets\cite{Kahn,Gatteschi94} has emerged in the 
last decade as one of the major interdisciplinary fields of research in 
the materials science community. Molecular magnets, also called ``spin 
clusters'', are crystalline materials composed of magnetic centers, 
mostly transition-metal ions, which have strong mutual interactions 
within each molecule. Each spin cluster in the lattice is magnetically 
well isolated from its neighbors due to the presence of surrounding
ligands. This magnetic shielding allows the study of the individual 
behavior of nanometer-scale magnetic systems.

\begin{figure}[t!]
\centering
\includegraphics[height=65mm,angle=0]{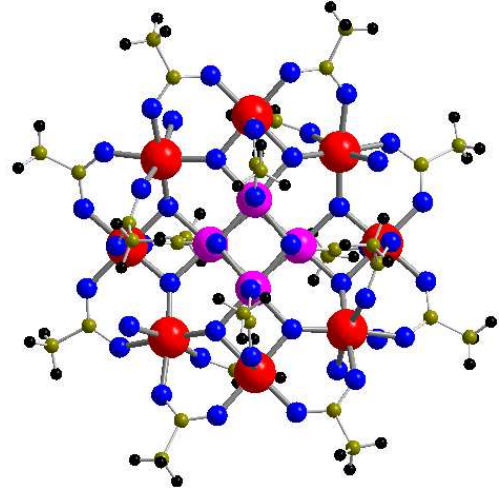}
\caption{(Color online) Structure of $\rm Mn_{12}$-acetate viewed along 
the $c$ axis. Large solid spheres represent $\rm Mn^{3+}$ ions (outer 
ring) and $\rm Mn^{4+}$ ions (inner core). All $\rm Mn$ atoms have a 
distorted octahedral coordination geometry.}
\label{fig_mn12}
\end{figure}

Chemists and physicists have combined efforts in an increasingly coordinated 
approach to develop and design new magnetic molecules which can show 
purely quantum properties at the macroscopic level. Such phenomena include
very slow relaxation of the magnetization below the ``blocking'' temperature 
$T_B$.\cite{Sessoli02,Barbara99} and quantum tunneling of the magnetization 
vector through an energy barrier between ``spin up'' and ``spin down'' 
configurations.\cite{QTref} The barrier $\Delta$ is governed, to leading 
order, by two parameters, the ground-state spin value $S$ and the axial 
anisotropy $D \leq 0$ of the cluster, which yield $\Delta = -D S^2$. The 
magnetic properties of the molecules are determined primarily by the exchange 
interactions between individual ionic spins and by their magneto-crystalline 
anisotropy. Competition among the different exchange couplings can 
lead to a wealth of situations ranging from a nonmagnetic ground 
state ($S = 0$) to a variety of large-spin ground states ($S \geq 10$). 
A precise knowledge of the values of the exchange interactions, and 
their dependence on the geometry of the molecular cluster, is therefore 
critical to understand the ground state and, more generally, the 
possibilities for new synthetic procedures providing better materials 
with larger energy barriers.

Inelastic neutron scattering (INS) is a powerful experimental probe 
of magnetic excitations and exchange or anisotropy parameters in 
molecular magnetic clusters.\cite{Hennion97,Mirebeau99,Andres00,
Amoretti00,Caciuffo98,Chaboussant02,Andres02,Basler02,Ochsenbein03,Basler03}
Here we investigate the high-energy magnetic states in the 
best-characterized molecular magnetic system to date, 
$\rm Mn_{12}$-acetate, using INS in the energy range between $1$ and 
$100$ meV (10-1000 K) and with no external magnetic field. Comparison 
between our experimental results and numerical calculations enables 
us to estimate the exchange interaction parameters in the 
$\rm Mn_{12}$-acetate cluster.

The manuscript is organized as follows. In Sec.~\ref{secSP} we review the 
structure and properties of $\rm Mn_{12}$-acetate, experimental results, 
and attempts which have been made to understand the exchange interactions. 
In Sec.~\ref{secExp} we present the results of INS measurements performed
on three different spectrometers, and use these to characterize the magnetic 
excitations. Section \ref{secModAn} contains a theoretical analysis of the 
conventional exchange model for $\rm Mn_{12}$-acetate by high-temperature 
series expansion techniques for the susceptibility and exact diagonalization 
to obtain the ground state and low-lying excited states, both refined by 
comparison with the existing data. A summary and conclusions are presented 
in Sec.~\ref{secSum}.

\section{Structure and Properties of $\rm \bf Mn_{12}$-acetate}

\label{secSP}

\subsection{Structural Information}

\label{secSinf}

\begin{figure}[t!]
\includegraphics[height=45mm,angle=0]{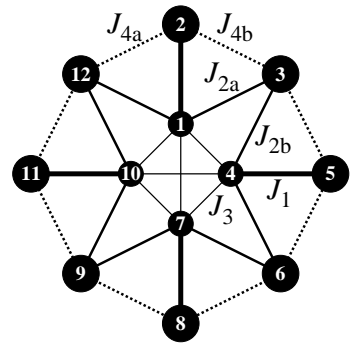}
\caption{Schematic representation of the magnetic exchange paths in $\rm 
Mn_{12}$-acetate as first proposed by Sessoli {\it et al.}\cite{Sessoli93a} 
$J_{1}$ couples $\rm Mn^{3+}$ and $\rm Mn^{4+}$ ions ($d = 2.771$ 
\AA), $J_{2a,b}$ correponds to $\rm Mn^{3+} / Mn^{4+}$ ($d = 3.45$ 
\AA), $J_{3}$ to $\rm Mn^{4+} / Mn^{4+}$ ($d = 2.817$ \AA) 
and $J_{4a,b}$ to nearest-neighbor $\rm Mn^{3+}$ ions 
($d = 3.33$ and 3.41 \AA).} \label{fig_exch}
\end{figure}

$\rm Mn_{12}$-acetate\cite{Sessoli93b,Sessoli93a,Caneschi99,Barbara99} 
is a mixed-valence ($\rm Mn^{3+}/Mn^{4+}$) compound where the magnetic 
ions are arranged in two groups: a central core composed of a tetrahedron 
of 4 $\rm Mn^{4+}$ ions ($S = 3/2$) and an external ring, or crown, of 8 
$\rm Mn^{3+}$ ions ($S = 2$). Figure \ref{fig_mn12} shows the $\rm 
Mn_{12}$-acetate cluster viewed along the $c$ axis. The point group of 
the Mn$_{12}$ molecule in the crystal structure is $S_4$. To simplify 
the analysis we make the additional assumption of fourfold rotation and 
reflection symmetry of the individual molecular clusters, and return later 
to a discussion of this approximation. Each cluster is only very weakly 
coupled to its neighbors, which are separated from each other by molecules 
of water and acetic acid, such that no long-range magnetic order has been 
found for temperatures as low as the mK range. Consequently, most of the 
experimental work performed at temperatures exceeding 1 K may be interpreted 
in terms of the properties of a single molecule. Neighboring $\rm Mn$ ions 
within a cluster are coupled in an intricate pattern by different types of 
$\mu$-oxo bridge and by acetate bridges, as a result of which both 
antiferromagnetic (AFM) and ferromagnetic (FM) exchange interactions 
may be present in the system. A schematic representation of the exchange 
couplings is shown in Fig.~\ref{fig_exch}. Within the approximation of 
fourfold cluster symmetry there are three inequivalent $\rm Mn$ sites 
with four different exchange couplings between neighboring $\rm Mn$ ions: 
$J_{1}$ (involving two $\mu$-oxo bridges) and $J_{2} = J_{2a} \approx 
J_{2b}$ (one $\mu$-oxo bridge) between $\rm Mn^{3+}$ and $\rm Mn^{4+}$ 
ions, $J_{3}$ between $\rm Mn^{4+}$ ions inside the core tetrahedron 
(two $\mu$-oxo bridges) and $J_{4} = J_{4a} \approx J_{4b}$ between 
$\rm Mn^{3+}$ ions around the external ring (one $\mu$-oxo bridge and 
two carboxylate groups). Inspection of the $\rm Mn$-$\rm Mn$ distances 
and $\rm Mn$-$\rm O$-$\rm Mn$ angles presented in Table \ref{table1} 
suggests that the approximations $J_{2a} \approx J_{2b}$ and $J_{4a} 
\approx J_{4b}$ are eminently reasonable.

\begin{table}[b!]
\caption{$\rm Mn$-$\rm Mn$ distances and $\mu$-oxo bridge angles for the 
different exchange couplings in Mn$_{12}$-acetate. $\rm Mn$ pairs 
coupled by one $\mu$-oxo bridge have greater separation ($d \geq 3.3$ \AA) 
and a higher angle than those coupled by two $\mu$-oxo bridges ($d \leq 3.3$ 
\AA).}
\label{table1}
\vspace{0.5cm} 
\begin{tabular}
{|c|c|c|} \hline  $\;$ exchange $\;$ & Mn-Mn  & Mn-O-Mn  \\
 path & $\;$ distance in {\AA} $\;$  &  angle \\ \hline
 {$J_1$}  &  2.771 & $\;$ 95.74$^\circ$, 94.00$^\circ$ $\;$  \\
 {$J_{2a}$} &  3.449 & 133.0$^\circ$  \\
 {$J_{2b}$} &  3.459 & 132.0$^\circ$  \\
 {$J_3$}  &  2.817 & 95.2$^\circ$, 95.0$^\circ$  \\
 {$J_{4a}$} &  3.331 & 122.52$^\circ$  \\
 {$J_{4b}$} &  3.410 & 129.15$^\circ$  \\
\hline
\end{tabular}
\end{table}

\subsection{Ground State and Anisotropy}

\label{secGS}

The magnetic ground state has a total spin $S = 10$, as first suggested 
by Sessoli {\it et al.}\cite{Sessoli93b,Sessoli93a} and later confirmed 
by many experimental studies, including magnetization,\cite{Thomas96} 
nuclear magnetic resonance (NMR),\cite{Furukawa00} electron paramagnetic 
resonance (EPR),\cite{Barra97,Hill98} and INS\cite{Mirebeau99} 
measurements. The $S = 10$ ground state may be rationalized by considering 
a ferrimagnetic arrangement of 8 parallel spins $S = 2$ on the crown 
$\rm Mn^{3+}$ ions oriented antiparallel to the 4 aligned $S = 3/2$ spins 
on the core $\rm Mn^{4+}$ ions. 

While all of the $\rm Mn$ atoms have a distorted octahedral coordination 
geometry, the distortion is significantly stronger for the crown Mn$^{3+}$ 
ions. The strongly negative axial anisotropy parameter $D$ of the cluster, 
which is due primarily to the Jahn-Teller distortion (axial octahedral 
elongation) around the $\rm Mn^{3+}$ ions, lifts the degeneracy of the 
$S = 10$ ground-state multiplet, with the $M_{S} = \pm 10$ states lying 
lowest. The $M_{S} = \pm 10$ and $M_{S} = 0$ states are separated 
by an energy barrier $\Delta= - D S_{z}^{2}$ of approximately 65 K (5.5 
meV), which blocks the thermal relaxation in a typical magnetization 
relaxation experiment.\cite{Sessoli93b,Gomes98,Zhong99,Bokacheva00} 
Surprisingly, the energy barrier $\Delta$ in $\rm Mn_{12}$-acetate 
remains the highest known, despite intense efforts to synthesize new 
magnetic clusters.

Higher-order (rhombic and quartic) terms in the anisotropy are 
responsible for a mixing of wave functions between pairs of states of 
equal $|M_{S}|$ on either side of the barrier, opening a channel 
for tunneling through the barrier. This effect has been demonstrated 
and studied extensively in recent years by a variety of techniques, 
including low-temperature 
magnetometry,\cite{Thomas96,Barbara00,Chiorescu00,Zhong00} 
NMR,\cite{Furukawa00} and specific heat.\cite{Gomes98,Luis00} An 
excellent determination of 
the quartic terms has been achieved using high-field EPR\cite{Barra97} 
and INS.\cite{Mirebeau99} The recent discovery\cite{Cornia02,Hill02,Hill03}
of disorder in the acetic-acid and water molecules, which further lowers the 
site symmetry of some of the $\rm Mn_{12}$ clusters present in the lattice, 
has shed additional light on a long-standing controversy concerning the 
physical origin of the experimentally observed 
tunneling.\cite{Barco03,Mertes01,Hernandez02,Amigo02,Chudnovsky01}

With the primary aim of understanding the magnetic excitation spectrum,
in this study we will treat the anisotropy contributions of single Mn 
ions using only phenomenological axial anisotropy parameters $D_{S}^0$ 
of the cluster spin $S$ ($D$ above denotes the ground-state anisotropy 
parameter, denoted $D_{10}^0$ in Sec.~\ref{secAss}), and neglect all 
anisotropy terms of higher order than quadratic in single-ion or cluster 
spins. Because single-ion anisotropies are some two orders of magnitude 
smaller than the exchange couplings, and higher-order anisotropies smaller 
still, this treatment is expected to be acceptable. Nevertheless, 
approximations at the level of single-ion anisotropies do limit the 
accuracy with which experimental results may be fitted in a theoretical 
analysis.

\subsection{Magnetic Excitations}

\label{secME}

While the low-temperature properties of $\rm Mn_{12}$-acetate are well 
described by an isolated $S = 10$ ground state, to date very little is 
known about the higher excited states whose energies are determined 
directly by the exchange interactions ($J_1,J_2,J_3,J_4$). The first 
experimental information was provided by magnetization measurements at 
very high fields using an explosive compression technique which can 
access magnetic fields in excess of $H = 900$ 
T.\cite{Zvezdin98,Gatteschi_review02} A series of peaks is 
observed between 300 T to 600 T in the field derivative $dM/dH$, suggesting 
important steps in the magnetization curve which may be interpreted in 
terms of crossovers from $S = 10$ to $S = 11$, 12, $\dots$, up to the 
maximum possible spin value of $S = 22$. Further information on the 
interaction parameters may be obtained from fits to the high-temperature 
part of the zero-field susceptibility. However, because the Curie-Weiss 
regime is not established even at $300$ K,\cite{Sessoli93a} and 
significant loss of solvent occurs at and above $308$ K,
it is difficult to draw a reliable conclusion from such fits
without a theoretical treatment of the susceptibility which is valid
also for lower temperatures. Such an analysis has not been performed 
previously, and will be presented in Sec.~\ref{secModAn} below.

Based on susceptibility and magnetization data alone, the unambiguous 
determination of exchange constants of similar magnitude in such 
ferrimagnetic systems is impossible. INS is a unique 
spectroscopic technique which can gather information over a wide energy 
range, and is thus capable of identifying a number of energy levels 
sufficient to evaluate the exchange interactions in complex systems 
such as $\rm Mn_{12}$-acetate.

Hennion {\it et al.}\cite{Hennion97}, using an approximately $30\%$ 
deuterated $\rm Mn_{12}$-acetate sample, showed clearly for the first 
time that important higher-energy transitions were observable above the 
transitions within the ground-state multiplet. These authors reported a 
series of inelastic peaks between $4$ meV and $10$ meV (45--110 K) which 
were ascribed to transitions between the $S = 10$ ground state and $S = 
9$ excited states. In this study we confirm the INS peaks observed in 
Ref.~\onlinecite{Hennion97}, providing more detailed information concerning 
these, and extend our survey to higher energies. Comparison between 
the experimental data and the spectra obtained from exact numerical 
diagonalization of a four-parameter model for the magnetic cluster 
allows us to assign the observed energy levels and to determine a 
consistent set of values for the exchange interactions in 
$\rm Mn_{12}$-acetate.

\subsection{Estimates of Exchange Interactions} 

\label{secEEI}

The first theoretical attempts to determine the exchange interactions 
considered a simplified 
model\cite{Sessoli93a,Katsnelson99,Raedt03,Alsaquer00} predicated on 
the assumption of a single AFM exchange interaction $J_{1}$ much larger 
than any of the other parameters in the system. In this situation the 
4 $\rm Mn^{3+}$-$\rm Mn^{4+}$ pairs coupled by $J_{1}$ are locked in their 
$S = 1/2$ ground state, and the size of the spin space is considerably 
reduced, namely to 4 spins $S = 1/2$ and 4 spins $S = 2$. Although this 
model provided a qualitative account of the available susceptibility data, 
it was far from providing a unique answer. Further, from qualitative 
considerations of exchange interactions, and particularly the proximity 
of the $J_1$ $\mu$-oxo bridging angle to 90$^\circ$, even this simplifying 
assumption lacks justification. In short, a fully general starting point is 
required, and the absence of detailed information concerning the location 
of the magnetic excitations precludes any unambiguous determination of the 
exchange couplings.

Several attempts have been made to evaluate the exchange interactions 
by applying numerical techniques, specifically exact 
diagonalization\cite{Raghu01,Regnault02} and different variants of 
density functional theory (DFT) calculations,\cite{Boukhvalov02,Park03} 
to a model of the type discussed above. The results of these studies 
were then compared with a selectively chosen subset of the available 
data, taken from high-field magnetization,\cite{Zvezdin98,Gatteschi_review02}
a.c.~susceptibility,\cite{Mukhin98,Gomes98} high-field EPR,\cite{Edwards03} 
and INS\cite{Hennion97} measurements, to argue in support of a particular 
parameter set. Raghu {\it et al.}\cite{Raghu01} 
considered the parameters proposed in Ref.~\onlinecite{Sessoli93a}, 
and demonstrated that these do not in fact yield an $S = 10$ ground 
state. Variants on this set were tested which did satisfy this condition, 
but we will show below that the very large values of $J_1$ [215 K (18.5 
meV)] and of $J_2$ and $J_3$ [85 K (7.3 meV)] are inconsistent with the 
measured susceptibility. Regnault {\it et al.},\cite{Regnault02} focusing 
on the $S = 10$ condition, obtained $J_{1} = 119$ K (10.2 meV), $J_{2} = 
118$ K (10.2 meV), $J_{3} = -8$ K ($- 0.69$ meV), and $J_{4} = 23$ K (1.98 
meV), where the magnitudes of the parameters were determined by fitting the 
magnetization data. Park {\it et al.}\cite{Park03} obtained a not dissimilar 
result in an {\it ab initio} DFT calculation, specifically $J_{1} = 115$ K 
(9.91 meV), $J_{2} = 84$ K (7.24 meV), $J_{3} = - 4$ K (0.34 meV), and 
$J_{4} = 17$ K (1.47 meV). In both of these cases, $J_{1}$ and $J_{2}$ are 
rather close in value, in sharp contrast to the assumption underpinning the 
simplified, dimerized model. While $J_{1}$ is much smaller than previously 
suggested,\cite{Sessoli93a} in combination with the larger $J_2$ value 
these sets also overestimate the high-temperature susceptibility quite 
considerably. The much weaker $J_{3}$ and $J_{4}$ interactions are then 
difficult to determine systematically, although both sets suggest that 
$J_3$ may be weakly FM. Finally, Boukhvalov {\it et al.}\cite{Boukhvalov02} 
obtained a smaller set of values, $J_{1} = 47$ K (4.1 meV), $J_{2} = 26$ K 
(2.2 meV), two different core-spin coupling terms $J_{3} = 30$ K (2.6 meV) 
and $J_3' = 10$ K (0.86 meV), and $J_{4} = 7$ K (0.60 meV). However, from 
our computations these last values appear to give neither a ground state 
with $S = 10$ nor a suitable reproduction of the magnetization.

It is thus clear that no consensus has yet emerged on the exchange 
coupling values in $\rm Mn_{12}$-acetate. The objective of this study 
is to determine these exchange interactions based on the present INS 
results for the previously unexplored energy range up to 45 meV (500 K), 
in combination with information obtained from susceptibility and 
magnetization data. We will find that it is possible to determine an 
effectively unique set of interaction parameters capable of explaining 
consistently all of the available information, i.e.~the $S = 10$ ground 
state, the high-temperature susceptibility, the approximate location 
of the observed magnetization steps, and the magnetic excitations 
measured by INS.

\section{Experiment}

\label{secExp}

\subsection{Experimental Details}

The INS experiments were performed on three different instruments located 
at three different sources: 1) the time-of-flight (TOF) spectrometer IN4 at 
the Institut Laue-Langevin (Grenoble, France) using wavelengths $\lambda = 
1.1$ \AA\ and $2.2 $ \AA, 2) the TOF spectrometer FOCUS at the Paul Scherrer 
Institute (Villigen, Switzerland) using the wavelength $\lambda = 3.1$ 
\AA, and 3) the MARI spectrometer at the pulsed neutron source 
ISIS (Didcot, UK) using neutrons with incident energies between 12 meV 
and 50 meV (in this section we quote energies only in meV). IN4 and MARI 
are best suited for high energy transfers and a wide $Q$-range, while 
FOCUS is a cold-neutron spectrometer designed for energy transfers below 
5-10 meV. Data were collected at several temperatures between $1.5$ K 
and $100$ K, and corrected for the background and detector efficiency by 
means of a vanadium reference and empty-cell measurements for each incident 
energy.

On IN4 and FOCUS, we used a fresh, $3.8$ g polycrystalline powder 
sample of fully deuterated $\rm Mn_{12}$-acetate placed under 
helium in a flat, rectangular aluminum slab of $3$ mm thickness. 
Full deuteration was achieved by systematic substitution of 
hydrogenated precursors with deuterated versions under argon. On 
MARI, the sample, wrapped in aluminum foil, was mounted in a hollow 
cylinder of diameter $42$ mm and height 62 mm. The sample thickness 
was approximately 3 mm.

On IN4, the $\rm ^{3}He$ detector banks cover the angular range 
$2 \theta = 13$-$120^\circ$, giving access to momentum transfers $ 0.7 
\leq Q \leq 4$ {\AA}$^{-1}$ at $\lambda = 2.2$ {\AA} and $2 \leq Q 
\leq 9$ {\AA}$^{-1}$ at $\lambda = 1.1$ {\AA}. The resolution obtained 
from a metallic vanadium reference, given as the full-width at 
half-maximum peak height of the elastic line, was $\Gamma = 0.9$ meV 
at 2.2 {\AA} and $\Gamma = 3.8$ meV at 1.1 {\AA}. On FOCUS, the detector 
banks cover the angular range $2\theta = 10$-$130^\circ$, giving access 
to momentum transfers $0.7 \leq Q \leq 3.4$ {\AA}$^{-1}$ at $\lambda 
= 3.1$ {\AA}. FOCUS is a time- and space-focusing TOF spectrometer which 
allows inelastic focusing. At an energy loss of approximately $5$ meV, the 
resolution, obtained from a metallic vanadium reference, was $\Gamma = 
0.4$ meV at 3.1 {\AA}. On MARI, the detectors cover the angular range $2 
\theta = 12$-$135^\circ$ and momentum transfers up to $Q \approx 8$-9 
{\AA}$^{-1}$ for the highest energy transfer used (50 meV). The elastic 
resolution ranged from 0.15-0.2 meV at an incident energy $E_{i} = 
12$ meV to 1 meV at $E_{i} = 50$ meV.

\subsection{Magnetic Neutron Cross-Section}

\label{secINScr}

The differential magnetic neutron cross-section for a transition
$|\Psi_{m} \rangle \rightarrow |\Psi_{n} \rangle$ is\cite{Marshall71}
\begin{eqnarray}
\frac{d^2\sigma}{d\Omega dE} & = & \frac{N}{4}\left\{ {\frac{\gamma
e^2}{m_e c^2}} \right\}\frac{k'}{k} e^{-2W(\bm{Q})} F^{2}(\bm{Q}) 
\nonumber \\ & & \times \sum\limits_{\alpha,\beta} \left\{{\delta_{\alpha 
\beta} - \frac{Q_\alpha Q_\beta }{\bm{Q}^2}}\right\} \label{eq:INS} \\ 
\nonumber & & \times \sum\limits_{i,j} \exp^{i \vec{Q} \cdot (\vec{R}_{i} - 
\vec{R}_{j})}\left\langle {\Psi_m } \right| \hat {S}_i^\alpha \left| 
{\Psi_n } \right\rangle \\ \nonumber & & \times \left\langle {\Psi_n } 
\right|\hat {S}_j^\beta \left| {\Psi_m } \right\rangle \delta 
\left( {\hbar \omega + E_n - E_m } \right). 
\end{eqnarray}
In this equation $N$ is the number of Mn$_{12}$ molecules in the sample, 
$k$ and $k'$ are the wavenumbers of the incoming and scattered neutrons, 
$\bm{Q}$ is the scattering vector, $\exp(-2W(\bm{Q}))$ is the Debye-Waller 
factor, $\hbar \omega$ is the neutron energy, $|\Psi_m \rangle$ are the 
cluster wave functions with energies $E_{m}$, $F(\bm{Q})$ is the 
magnetic form factor, $\vec {R}_i$ is the space vector of the $i$th 
$\rm Mn$ ion in the cluster, $\alpha $ and $\beta $ represent the 
spatial coordinates $x$, $y$, and $z$, $e$ and $m_{e}$ are respectively 
the charge and mass of the electron, $c$ is the speed of light, and 
$\gamma  = - 1.91$ is the gyromagnetic constant of the neutron. INS 
selection rules impose that the matrix elements in Eq.~(\ref{eq:INS}) 
are nonzero only when $\Delta S = S - S' = 0, \pm 1$ and $\Delta M_{\rm 
S} = M_{S} - M'_{S} = 0, \pm 1$ where the initial and final 
states are defined respectively by $(S,M_{S})$ and $(S',M'_{S})$. 

\subsection{Energy-dependence}

\begin{figure}[t!]
\includegraphics[height=155mm,angle=0]{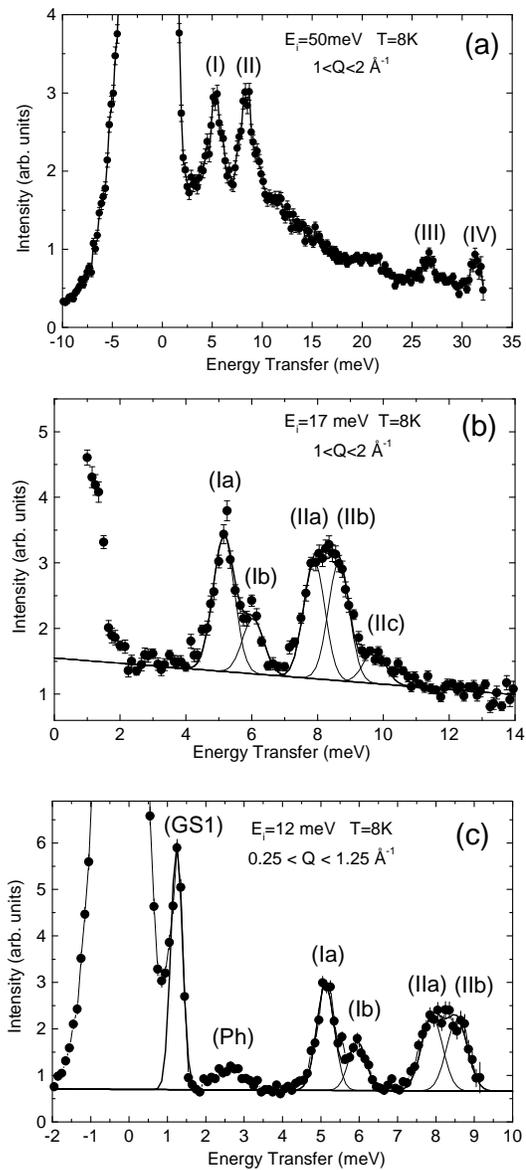}
\caption{INS spectra obtained at $T = 8$ K on MARI at (a) $E_{i} = 50$ 
meV, (b) $E_{i} = 17$ meV, and (c) $E_{i} = 12$ meV. The $Q$-range 
is restricted to $1 \leq Q \leq 2 $ {\AA}$^{-1}$ in panels (a) and (b) 
and to $0.25 \leq Q \leq 1.25 $ {\AA}$^{-1}$ for panel (c). Transitions 
are labeled as discussed in the text. In panel (c), (GS1) is a transition 
within the $S = 10$ ground-state multiplet and (Ph) denotes a low-energy 
phonon mode. }
\label{fig_mari}
\end{figure}

Figure \ref{fig_mari} shows inelastic spectra obtained on MARI at $T = 8$ 
K at three different incident energies ($E_{i} = 50, 17$, and $12$ meV). 
Several transitions are observed between 1.2 meV and 32 meV. At $E_{i} 
= 50$ meV [Fig.~\ref{fig_mari}(a)], four sets of transitions are visible 
above a broad background produced by incoherent scattering. These sets 
are labeled (I) to (IV), and their peak energies are $5.3$, $8.5$, $27$, 
and $31$ meV, respectively. The positions of (I) and (II) are consistent 
with the observations made by Hennion {\it et al.}\cite{Hennion97}

Figure \ref{fig_mari}(b), corresponding to $E_{i} = 17$ meV, shows 
that peaks I and II are composed of several components. These are 
labeled (Ia), (Ib), (IIa), (IIb), and (IIc), and from the center 
lines of Gaussian fitting curves appear respectively at energies 
of $5.15$, $6.04$, $7.90$, $8.64$, and $9.73$ meV. While a 
double-peak structure is clearly established for transitions (Ia) and 
(Ib), it is less obvious for the higher-energy transitions. The peaks 
were fitted using Gaussian lineshapes by assuming a constant line width 
$\Gamma \approx 0.75$ meV for all transitions and a linearly decreasing 
slope to account for the background. The width is determined from 
transitions (Ia) and (Ib) and its value is maintained for the other 
transitions. This is again in qualitative agreement with Fig.~8 of 
Ref.~\onlinecite{Hennion97}.

Figure \ref{fig_mari}(c) confirms clearly that there is an energy 
splitting between transitions (Ia) and (Ib) but transitions (IIa) and 
(IIb) remain (barely) unresolved. At this value of $E_i$ the width of 
these transitions is between $0.45$ and $0.55$ meV. Based on its $Q$-
and $T$-dependence (not shown), the broad peak around 2.65 meV may be 
attributed to phonon excitations. In addition, with this resolution it 
is now possible to observe the transition peak within the ground state, 
labeled (GS1), at $\hbar \omega \approx 1.25$ meV, in complete agreement 
with the results of Mirebeau {\it et al.}\cite{Mirebeau99} This peak 
is significantly sharper ($\Gamma \approx 0.35$ meV) than the 
higher-energy transitions, suggesting an intrinsic line-broadening of 
transitions (Ia) to (IIb).

\subsection{\mbox{$\bm{T}$}-dependence}

\begin{figure}[t!]
\includegraphics[height=110mm,angle=0]{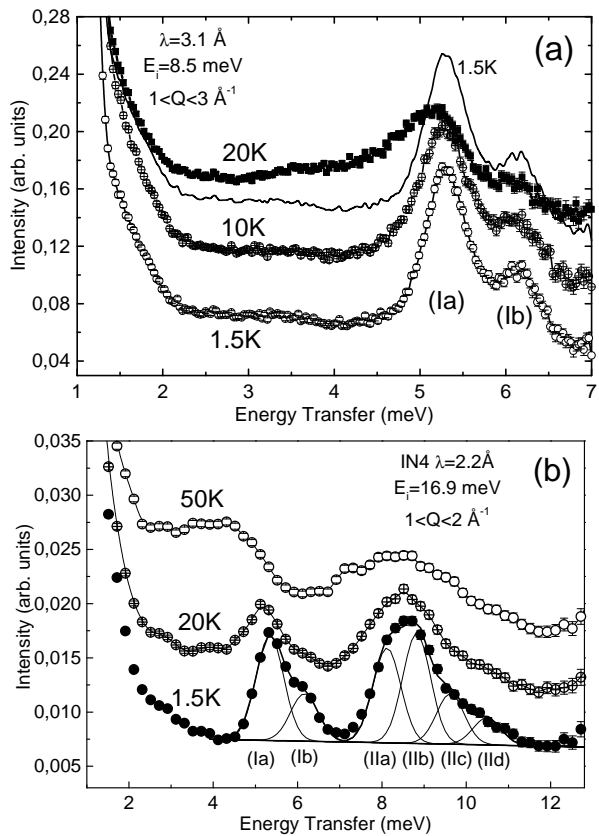}
\caption{(a) Inelastic spectra at $T = 1.5$ K, $10$ K, and $20$ K 
obtained on FOCUS at an incident wavelength $\lambda = 3.1$ {\AA}. 
At $1.5$ K, transitions (Ia) and (Ib) are observed at $\hbar \omega = 
5.3$ meV and $\hbar \omega = 6.2$ meV, respectively. The data taken at 
1.5 K are reproduced as a solid line and shifted upwards to facilitate 
comparison with the 20 K data.(b) Inelastic spectra at $T = 1.5$ K, 
$20$ K, and $50$ K obtained on IN4 at $\lambda = 2.2$ {\AA}. } 
\label{figTdep}
\end{figure}

Figure \ref{figTdep}(a) shows the inelastic spectra obtained on FOCUS at 
temperatures of $1.5$ K, $10$ K, and $20$ K, after integrating over a 
group of detectors such that the $Q$-range is $1 \leq Q \leq 3 $ 
{\AA}$^{-1}$. A vertical offset is applied between the curves for 
clarity. The results provide a significant improvement in the 
understanding of transitions (Ia) and (Ib): these have energies of $\hbar 
\omega_{Ia} \approx 5.3$ meV and $\hbar \omega_{Ib} \approx 6.2$ meV at 
1.5 K, consistent with the MARI data (taken at 8 K). As the temperature 
is increased there is a shift of both peaks towards lower energies, and 
some broad scattering develops below 5 meV. This is shown clearly 
by comparison of the data taken at $20$ K and $1.5$ K (solid line), 
for which the same vertical offset is applied. The shift to lower 
energies arises from progressive population of the higher $M_{S}$ 
sublevels in the $S = 10$ ground-state multiplet, which results in a 
large number of new magnetic transitions visible by INS and permits 
an estimate of the anisotropy splitting in the excited states (see 
Sec.~\ref{secAss}).

Figure \ref{figTdep}(b) presents inelastic spectra obtained on IN4 over 
a range of temperatures between $1.5$ K and $50$ K, where the $Q$-range 
is limited to $1 \leq Q \leq 2 $ {\AA}$^{-1}$. At $1.5$ K one finds the 
transitions discussed above, which weaken progressively as the temperature 
increases, and finally merge into a single, broad feature in which both 
magnetic and phonon scattering are present. A general shift of intensity to 
lower energies is observed with increasing temperature for both peaks groups 
I and II. For the low-temperature data, by keeping the line width fixed 
to the value obtained by fitting transitions (Ia) and (Ib), $\Gamma = 0.77$ 
meV, the best fit to the large band at higher energies is obtained by 
considering four peaks, (IIa) to (IId), with energies of $8.1$, $8.8$, 
$9.6$, and $10.5$ meV. The first three peaks are globally consistent with 
the data from MARI [Fig.~\ref{fig_mari}(b)]. However, the final peak, 
(IId), is very weak and does not seem to be present in the MARI data, 
so its existence must be said to be questionable. 

\subsection{\mbox{$\bm{Q}$}-dependence}

\begin{figure}[t!]
\includegraphics[height=50mm,angle=0]{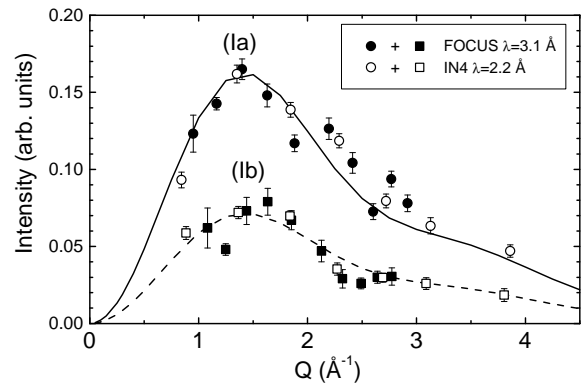}
\caption{$Q$-dependence at $1.5$ K of peaks (Ia) and (Ib) obtained 
from the inelastic spectra shown in Fig.~\ref{figTdep}(a). Data from 
both FOCUS and IN4 are shown after normalization of the maximum intensity 
at $Q \approx 1.45$ {\AA}$^{-1}$.} \label{figQplot}
\end{figure}

We consider next the $Q$-dependence of transitions (Ia) and (Ib). The 
integrated intensity is shown in Fig.~\ref{figQplot} for the two 
transitions, with the FOCUS data represented by solid circles and 
squares, and the equivalent IN4 data by open symbols. Each integrated 
intensity point was obtained by fitting the peaks using a Gaussian 
function whose center, width, and intensity were all allowed to float.

For both peaks, the intensity passes through a maximum at $Q 
\approx 1.45$ {\AA}$^{-1}$, and decreases progressively at higher 
$Q$ values. This $Q$-dependence is typical of magnetic scattering, 
where the intensity is expected to follow the magnetic form factor 
$|F(Q)|^{2}$, a rapidly decreasing function above $Q \approx 2$-3 
{\AA}$^{-1}$, modulated by a structure form factor $I(Q)$ which 
depends on the exchange couplings and the exchange connectivity 
within the cluster.\cite{Furrer79} In the simplest exchange-coupled 
system, a dimer, the structure form factor is given by\cite{Furrer79}
\begin{equation}
I(Q)  \sim \left( 1 - \frac{\sin(QR)}{QR} \right),
\end{equation}
where $R$ is the inter-atomic distance in the dimer. Using the 
dimer expression as an approximate indication, the maximum observed 
at $Q \approx 1.45$ {\AA}$^{-1}$ corresponds to $R \approx 2.8$ 
{\AA}, a value close to the shortest $\rm Mn$-$\rm Mn$ separation 
in $\rm Mn_{12}$-acetate, $d = 2.77$ \AA. By comparison, the 
$Q$-dependence of the ground-state transition at 1.25 meV is peaked 
at $Q \approx 0.95$ \AA$^{-1}$,\cite{Hennion97} consistent with 
its different physical origin.

\begin{figure}[t!]
\includegraphics[height=115mm,angle=0]{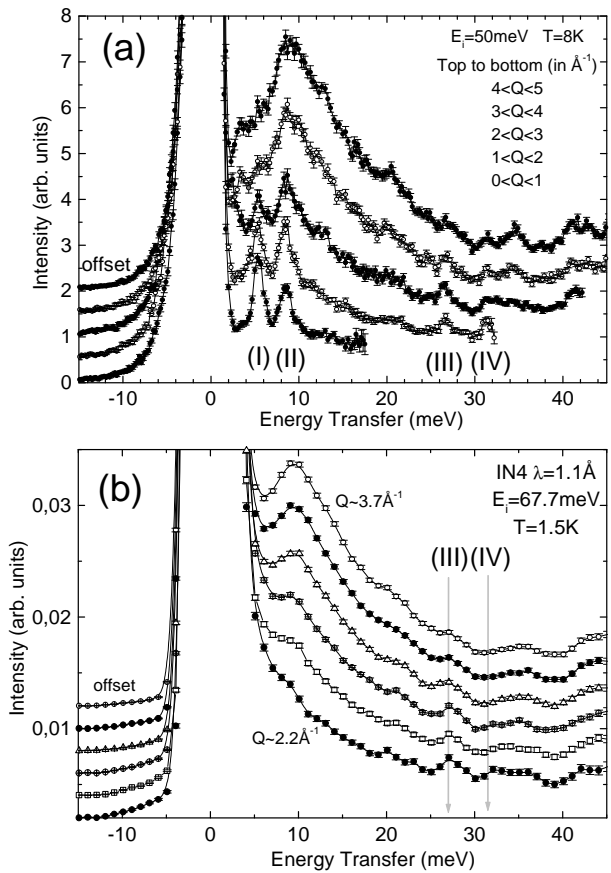}
\caption{(a) INS spectra obtained on MARI at $E_{i} = 12$ meV and 
$T = 8$ K for different values of $Q$. (b) INS spectra from IN4 at 
$\lambda = 2.2$ \AA\ and $T = 1.5$ K for different values of $Q$.} 
\label{fig_highQ}
\end{figure}

Figures \ref{fig_highQ}(a) and (b) show the behavior of the INS 
spectra obtained respectively on MARI and IN4 over the full $Q$-range 
of each instrument. In both cases, the low-energy transitions (I) and 
(II) discussed above are lost rapidly in the incoherent phonon 
background as $Q$ increases. The $Q$-dependence of peaks (III) and (IV) 
is more difficult to determine due to their weak intensity, but it is 
consistent with transitions of magnetic origin in that the intensity 
decreases with increasing $Q$. It is clear from both panels of 
Fig.~\ref{fig_highQ} that the 27 meV peak is more pronounced at small 
$Q$ values. A general increase of scattering is observed as $Q$ is 
increased for all energies between 10 and 45 meV, which unfortunately 
impedes a complete analysis.

\subsection{Assignment of Magnetic Peaks}

\label{secAss}

Several magnetic peaks have been observed in four groups with energies 
$\hbar \omega_{l}$ as shown in Table \ref{tabSum}. For transitions (Ia) and
(Ib), a detailed analysis may be performed of their dependence on $Q$ and on 
temperature; for the higher transitions, only a more qualitative 
treatment is possible with the present data.

To determine the nature of these transitions, we begin by noting that 
due to the negative axial zero-field splitting only the $M_{S} = 10$ 
component of the $S = 10$ ground state is populated (over $99.5\%$) at 
1.5 K, and therefore that only INS transitions from the state $(10,\pm 10)$ 
can be observed. From the selection rules in Sec.~\ref{secINScr}, it follows 
that only states with $S = 9,10$, and 11 can be excited by INS, a clear 
example of which is the observation\cite{Mirebeau99} of the transition from 
$E(10,\pm 10)$ to $E(10,\pm 9)$ within the $S$ = 10 ground-state multiplet 
at $\hbar \omega \approx 1.25$ meV [peak (GS1) in Fig.~\ref{fig_mari}(c)]. 
This result leads to the accurate determination of the axial anisotropy 
parameter for the $S = 10$ ground state, $D_{10}^0 = -0.0566$ meV.

\begin{table}[b!]
\caption{Summary of energies and spin states of the 
observed magnetic transitions.} \label{table_summary}
\vspace{0.5cm} 
\begin{tabular}
{|c|c|c|} \hline $l$ & $\hbar \omega_{l}$ & Properties \\ \hline
Ia & $(5.3 \pm 0.1) $ meV & $S = 9, M_{S} = \pm 9$ \\
Ib & $(6.2 \pm 0.1) $ meV & $S = 9, M_{S} = \pm 9$ \\
IIa & $(8.0 \pm 0.1) $ meV & $S = 9, M_{S} = \pm 9$ \\
IIb & $(8.7 \pm 0.1) $ meV & $S = 9, M_{S} = \pm 9$ \\
IIc & $(9.7 \pm 0.1) $ meV & $S = 9, M_{S} = \pm 9$ \\
IId & $\;\;$ $(10.5 \pm 0.1) $ meV $\;\;$ & not confirmed \\
$\;\;$ III $\;\;$ & $(27 \pm 1) $ meV & $S = 9$, $10$, or $11$ \\
IV & $(31 \pm 1) $ meV & $\;\;$ may be magnetic $\;\;$\\ \hline
\end{tabular}
\label{tabSum}
\end{table}

The assignment of spin states for the excited energy levels depends both 
on experimental observation and on certain assumptions concerning the 
axial anisotropy parameters for these states. We present first our 
treatment of anisotropy parameters, followed by the experimental results 
and a discussion of the consistency of this approach, which leads to the 
spin assignments shown in Table \ref{tabSum}. As noted in Sec.~\ref{secGS}
and discussed in greater detail in Sec.~\ref{secModAn}, given the nature 
of the data and the energy scales involved we restrict our considerations 
to the level of anisotropy parameters of the total spin. The canonical 
energy-level splitting for a given $S$ state may be expressed as
\begin{equation}
E(S,M_{S}) = D_{S} \left( M_{S}^{2} - {\textstyle \frac{1}{3}} S(S + 1) 
\right), \label{eaas}
\end{equation}
where $D_{S}$ is the axial anisotropy parameter of the cluster in 
spin state $S$. We will not attempt to compute the anisotropy parameters 
$D_{S}$ from vector coupling of the individual single-ion anisotropy 
terms, but for a qualitative interpretation of our observations will use 
related considerations to deduce their values relative to the measured 
ground-state axial anisotropy parameter $D_{10}^0$. In the absence of 
experimental information on this point, we begin by assuming that 
$D_{S}$ is the same for all excited states of the same spin $S$, and 
return below to a more detailed consideration of this approximation. 

A qualitative picture of the relative values of $D_{S}^0$ for the 
lowest-energy states of each spin $S$ in the Mn$_{12}$-acetate cluster 
may be obtained from a crude model based on two observations. From our 
approximation to the crystallographic symmetry all single-ion anisotropy 
terms are identical for both types of ion, and from the Jahn-Teller 
distortions these are very much greater for the 8 Mn$^{3+}$ sites than 
for the 4 Mn$^{4+}$ sites. A vector-coupling calculation of the spin 
state anisotropy performed only for the Mn$^{3+}$ crown sites gives a 
systematic decrease of $|D_{S}^0|$ with decreasing $S$, and no change 
in the sign. In this elementary approach the anisotropy parameter deduced 
for $S$ = 9 states, assumed to be composed of 3 Mn$^{3+}$-Mn$^{3+}$ pairs 
with $S$ = 4, one such pair with $S$ = 3, and an antiparallel $S$ = 6 
core, is approximately 8\% smaller than that for the $S$ = 10 ground 
state. From a comparison with the coefficient $D_{10}^0 = -0.0566$ meV 
extracted from INS data in Ref.~\onlinecite{Mirebeau99}, this result is 
fully consistent with the value $D_{9}^0 \approx - 0.049$ meV estimated 
by fitting the magnetization to a similar dimerized model.\cite{Tupitsyn00} 

For the uniaxial anisotropy parameter of the lowest excited $S$ = 11 
states, the simplest argument is to view these as a fully aligned $S$ = 
16 crown of Mn$^{3+}$ ions, with changes of spin state from the $S$ = 10 
ground state occuring on the core Mn$^{4+}$ ions, which have a very small 
single-ion anisotropy, whence $D_{11}^0 \approx D_{10}^0$. The vector-coupling 
scheme above gives a value for $D_{11}^0$ somewhat larger than $D_{10}^0$, 
but there is no qualitatively significant change in size or sign. The most 
important difference between $S$ = 9 and $S$ = 11 excited states for our 
purposes is that the latter should show three excitation branches 
corresponding to the $(S,M_{S})$ final states $(11,\pm 11)$, 
$(11,\pm 10)$, and $(11,\pm 9)$. These should be similar in intensity 
(differing by less than one order of magnitude) and should have a total 
energy span $\Delta E_{11} = 40 D_{11}^0$ [see Eq.~(\ref{eaas})].

We now demonstrate that the temperature-dependence of magnetic peaks 
(I) is consistent with transitions from the ground state $(10,\pm 10)$ to 
excited states with $S = 9$ and $M_{S} = \pm 9$. The $Q$-dependence 
of the magnetic transitions at $\hbar \omega_{Ia} = 5.3$ meV and $\hbar 
\omega_{Ib} = 6.2$ meV is shown in Fig.~\ref{figQplot}. The decrease of 
intensity at low $Q$ values is typical of transitions with $\Delta S = 
\pm 1$, which have zero intensity at $Q = 0$.\cite{Furrer79} In 
contrast, transitions with $\Delta S = 0$ are strong at $Q = 0$. 
Further, as the temperature increases (Fig.~\ref{figTdep}) there is 
a growing scattering intensity below $\hbar \omega_{Ia}$. This suggests 
that additional magnetic transitions are emerging because the increasing 
temperature causes thermal population of more $M_{S}$ sublevels of the 
$S = 10$ ground-state multiplet. From the value and sign expected for the 
parameter $D_9^0$, such a situation is possible only if the excited states 
corresponding to transitions (Ia) and (Ib) are two separate $S = 9$ states. 

\begin{figure}[t!]
\includegraphics[height=100mm,angle=0]{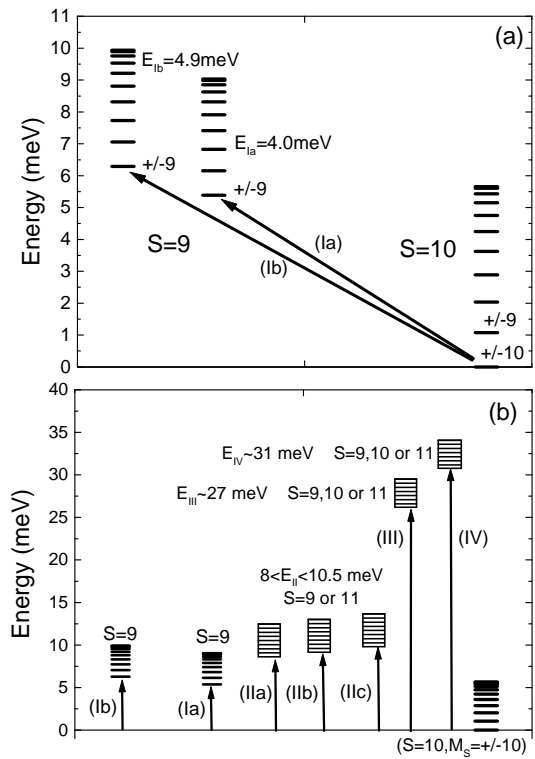}
\caption{(a) Detail of the low-energy levels inferred from our INS data 
(see text). After subtraction of the anisotropy shift, the lowest $S = 9$ 
states appear at $4.0$ meV and $4.9$ meV above the $S = 10$ ground state. 
(b) Energy levels observed by INS over the full range investigated. 
The spin values and anisotropy parameters of the higher levels are not 
known precisely.}
\label{fig_diagram}
\end{figure}

By applying Eq.~(\ref{eaas}), the energies of transitions from the ground 
state $(10,\pm 10)$ to $S = 9$ spin states with $M_{S} = \pm 9$ can 
be written in the form
\begin{eqnarray} 
\hbar \omega_{\alpha} & = & E(9,\pm 9) - E(10,\pm 10) \nonumber \\ 
& = & E_{\alpha} + {\textstyle \frac{190}{3}} |D_{10}^0| - 51 |D_{9}^0|,
\end{eqnarray}
where $E_{\alpha}$ is the energy of the $S = 9$ state in the 
absence of anisotropy. The multiplet structures are illustrated in 
Fig.~\ref{fig_diagram}(a), which shows also that the energy spread within 
the ground-state multiplet ($100D_{10}^0$) is larger than that within the 
$S = 9$ manifolds ($81D_{9}^0$). As temperature increases, the state 
$(10,\pm 9)$ and other, higher states within the $S = 10$ multiplet 
are populated progressively and give rise to more INS transitions, which 
will necessarily occur at lower energies. 

The measured changes in intensity can be fitted with semi-quantitative 
accuracy using the value $D_{9}^0 = - 0.045$ meV, and in fact our data 
suggest that the result of Ref.~\onlinecite{Tupitsyn00} should be regarded 
as an upper bound. With these cluster anisotropy parameters, the energetic 
shift induced in the transitions between the lowest $S = 10$ and $S = 9$ 
states is then
\begin{equation}
\hbar \omega_{\alpha} \approx E_{\alpha} + 1.29 \; {\rm meV}.
\label{eea}
\end{equation}
By contrast, if the excited states were of spin $S = 11$, for a reasonable 
value of $D_{11}^0$ one would expect the opposite process (stronger scattering 
at higher energies), which is clearly precluded by the present data. Further, 
an explanation based on one or more $S$ = 11 state(s) would require either a 
third excitation peak or that $D_{11}^0$ be very much smaller than $D_{10}^0$ 
so that all of the three peaks could be contained within the resolution width 
of the measurement; both of these possibilities may be excluded. 

The situation is less clear regarding transitions (IIa-d) between 8 and 
10.5 meV: it is evident from their evolution with temperature and with 
$Q$ that these are magnetic excitations. The intensity peaks undergo a 
small shift towards lower energies as the temperature is increased 
[Fig.~\ref{figTdep}(b)], while their wavevector-dependence (not shown) 
appears to have a weak maximum at intermediate $Q$. In principle these 
transitions could correspond to final states with $S = 9, 10$ or $11$. 
However, the combined $T$- and $Q$-dependence suggests that an $S$ = 10 
state is unlikely. In distinguishing between $S$ = 9 and 11 states one 
may invoke both the weak $T$-dependence in combination with expected
values for $D_{11}^0$, and the fact that the magnetization data of 
Refs.~\onlinecite{Zvezdin98,Gatteschi_review02} do not appear to show a
transition to an $S = 11$ state below a field of 200 T, implying that the 
first candidate $S = 11$ peak in the INS data would be (III). Thus from 
the qualitative similarity in behavior of peaks (II) with peaks (I), 
we assign peaks (II) to transitions with $S$ = 9 final states. In 
Sec.~\ref{secModAn} we will show that a numerical analysis of the 
Mn$_{12}$-acetate exchange model confirms this assignment, and can be 
used to exclude the possibility of a set of $S$ = 11 transitions with 
$D_{11}^0$ significantly different from $D_{10}^0$.

We comment briefly on the assumption that the cluster anisotropy parameters 
$D_{S}$ have the same value for all states of the same total spin $S$. We 
have used a simple model based on the classical spin arrangement of the 
$S = 10$ ground state and the strong difference in single-ion anisotropies 
between Mn$^{3+}$ and Mn$^{4+}$ ions to offer a consistent interpretation 
of all of the measured INS data. A small extension of this scenario is to 
consider the cluster as 4 Mn$^{3+}$-Mn$^{3+}$-Mn$^{4+}$ trimer units, 
constructing the ground state from 4 $S = 5/2$ trimers and excited states 
as perturbations about this. Under the same assumption concerning single-ion 
anisotropies, one obtains a range of values of $D_9$ for different $S = 9$ 
states from 5\% to 100\% of $D_{10}^0$; for excited $S = 10$ and $S = 11$ 
states, the values of $D_{10}$ and $D_{11}$ range from 40\% to 80\% of 
$D_{10}^0$. While this analysis is far from systematic, it illustrates three 
important qualitative features. First, it is difficult to find states with 
$D > D_{10}^0$, a result which is not surprising given the fully aligned 
crown spins in this state, and helps to verify the assumptions underlying 
the interpretation. Second, the sign of $D$ is always the same, suggesting 
that the temperature-dependence of the observed peaks can be used as a 
reliable indicator of the magnitude of $D_{S}$. Finally, this 
magnitude may in fact depend rather strongly on $S$; however, the 
constant sign is sufficient to ensure that the conclusions we have 
drawn from our INS data are robust against such quantitative changes. 

In conclusion, the $Q$-dependence of transitions (Ia) and (Ib) 
corresponds to transitions from the $S = 10$ ground state to $S = 9$ 
final states, and not to $S = 11$ or other $S = 10$ states. While a 
similarly quantitative analysis is not possible for the transitions 
(IIa-d), we have obtained sufficient evidence to conclude that these 
are also of $S = 9$ character. For transitions (III) and (IV) a 
detailed investigation is excluded, and, on the basis of the general 
intensity decrease with increasing $Q$, we state only that these are 
of magnetic origin. A summary of the magnetic excitations identified 
by INS, their energies, and their $S$ values is presented in Table 
\ref{table_summary}, and a graphical representation is provided in 
Fig.~\ref{fig_diagram}(b).

\section{Model Analysis}

\label{secModAn}

The microscopic Hamiltonian for exchange interactions between individual 
Mn ions in the Mn$_{12}$-acetate cluster (Fig.~2) may be expressed as 
\begin{eqnarray}
H & = & \sum_{j=1}^4 \{ J_1 \, \vec{S}_{3 j - 2} \cdot \vec{S}_{3 j - 1} 
+ J_3 \, \vec{S}_{3 j - 2} \cdot \vec{S}_{3 j + 1} \label{esh} 
\nonumber \\ & & \;\;\;\;\;\; + ( J_2 \, \vec{S}_{3 j - 2} + J_4 \, 
\vec{S}_{3 j - 1} ) \cdot ( \vec{S}_{3 j - 3} + \vec{S}_{3 j} ) \} 
\nonumber \\ & & \;\;\;\;\;\; + J_3 ( \vec{S}_1 \cdot \vec{S}_7 
+ \vec{S}_4 \cdot \vec{S}_{10} ), 
\end{eqnarray}
with periodic boundary conditions $i \equiv i + 12$. 
Sites $i$ = 1, 4, 7, and 10 represent the core $S = 3/2$ spins on the 
Mn$^{4+}$ ions, with mutual interactions $J_3$, and the other sites 
correspond to the $S = 2$ crown spins (Mn$^{3+}$), which have exchange 
interactions $J_1$ and $J_2$ with different core spins and $J_4$ with each
other around the outer ring (Fig.~\ref{fig_exch}). As stated in
Sec.~\ref{secSinf}, we neglect other possible exchange paths, and assume
despite the small structural distortions that the system is close to full
fourfold rotation and reflection symmetry, and thus that $J_{2a} = J_{2b} 
= J_2$ and $J_{4a} = J_{4b} = J_4$. We will return later to a quantitative 
statement on this latter point. For the numerical analysis to follow we will 
not consider single-ion anisotropy terms around individual ions, which are 
instead included through the term in Eq.~(\ref{eaas}) for the spin states of 
the cluster. We will also discard any of the possible higher-order anisotropy 
terms, such as those of the form $B_4^0 \, S_z^4$ and ${\textstyle 
\frac{1}{2}} B_4^4 \left(S_{+}^4 + S_{-}^4 \right)$. These approximations 
may be justified on the grounds of the small energy scale of the terms 
involved in comparison with the scale of the magnetic excitation spectrum 
(Sec.~\ref{secExp}). 

The aim of this section is to determine a set of exchange constants $\{J_i\}$
which is capable of explaining the magnetic excitation spectrum measured 
in Sec.~\ref{secExp}. Spectra of low-lying magnetic excitations have 
been tabulated by Raghu {\it et al.}\cite{Raghu01} by considering a 
variety of parameter sets. It is evident from this analysis that none of 
the sets considered yields sufficiently many excitations with appropriate 
total spin below an energy of approximately 300 K (30 meV) to explain the 
INS results. Similarly, the excitation spectra computed for the parameter 
set of Ref.~\onlinecite{Regnault02} also fails to provide a sufficient 
number of low-lying excitations to account for the new INS data (although 
we note here that these contain rather more levels in the relevant energy 
range than previous authors were aware of). While the parameter set of 
Ref.~\onlinecite{Park03} is not inconsistent with the excitation spectrum, 
we will find in Sec.~\ref{secSusc} that it cannot account for the 
magnetic susceptibility data. Hence the parameter sets discussed in 
Sec.~\ref{secEEI} are all incompatible with the experimental data. We 
therefore begin with an unbiased determination of an appropriate set of 
exchange constants $\{J_i\}$ by considering the ground state of the system 
using the full available parameter space. Candidate parameter sets are 
tested by computing the magnetic susceptibility, which we compare with 
our own measurements in Sec.~\ref{secSusc}. In Sec.~\ref{secHF} we comment 
briely on the high-field magnetization data. We return to the issue of 
magnetic excitations in Secs.~\ref{secAspec} and \ref{secSW}, where we 
present a theoretical analysis of the energy spectrum of the cluster.

\subsection{Numerical Methods}

The numerical results are obtained by two methods. We have performed a 
systematic high-temperature series expansion to compute the magnetic 
susceptibility
\begin{equation}
T \chi(T) = C_0 + C_1 \beta + C_2 \beta^2 + C_3 \beta^3 + C_4 \beta^4 + 
\ldots, \label{ehte2}
\end{equation}
where $\beta = 1/k_{\rm B} T$ with $k_{\rm B}$ the Boltzmann constant. 
For illustration, the first four coefficients $C_n$ for the cluster 
geometry and ionic spins of Mn$_{12}$-acetate are given by 
\begin{eqnarray}
C_0 & = & 21, \label{c} \nonumber \\
C_1 & = & -20 J_1 - 40 J_2 - {\textstyle \frac{75}{4}} J_3 - 64 J_4,
\nonumber \\
C_2 & = & -5 J_1^2 + 80 J_1 J_2 + 55 J_2^2 + 75 J_1 J_3 +
   150 J_2 J_3
\nonumber\\ &&
   + {\textstyle \frac{675}{16}} J_3^2 + 80 J_1 J_4 +
   160 J_2 J_4 + 112 J_4^2,
\nonumber \\
C_3 & = & {\textstyle \frac{211}{6}} J_1^3 + 20 J_1^2 J_2 - 80 J_1 J_2^2 +
   {\textstyle \frac{17}{6}} J_2^3 - {\textstyle \frac{225}{4}} J_1^2 J_3 
\nonumber \\ && 
    - 300 J_1 J_2 J_3 - 325 J_2^2 J_3 - {\textstyle \frac{675}{4}} J_1 J_3^2 
    - {\textstyle \frac{675}{2}} J_2 J_3^2 
\nonumber \\ && 
   - {\textstyle \frac{405}{64}} J_3^3 + 20 J_1^2 J_4 -
   380 J_1 J_2 J_4 - 280 J_2^2 J_4 
\nonumber \\ && 
   - 300 J_1 J_3 J_4 - 600 J_2 J_3 J_4 - 140 J_1 J_4^2 
\nonumber \\ && 
   - 280 J_2 J_4^2 - {\textstyle \frac{256}{15}} J_4^3. \label{eqCval}
\end{eqnarray}
One observes that $C_0$ is a constant, $C_1$ a weighted sum of the cluster 
interactions, $C_2$ a sum of certain combinations of squares, and so on. 
We have continued the series to 8th order, where the expressions are 
best handled by symbolic computation methods,\cite{Fukushima04} and 
have included the full Mn$_{12}$ cluster rather than employing a 
conventional linked-cluster expansion.

The second method we employ is the calculation of low-lying excitations 
for a given set of input interaction parameters $\{J_i\}$ by exact 
diagonalization (ED) of the cluster Hamiltonian (\ref{esh}) using the 
Lanczos procedure. We have used the conservation of $S_z$, the $z$-component 
of the total spin, as well as the spatial symmetries of the cluster. 
Conservation of $S_z$ is particularly important in reducing the dimension 
of the Hamiltonian matrix and is one of the reasons for which we do not 
include single-ion anisotropy terms in Eq.~(\ref{esh}). Spatial symmetries 
of the states are denoted by a ``momentum'' $k$ such that a state acquires a 
factor $\exp(ik)$ under a 90$^\circ$ rotation of the model for the Mn$_{12}$ 
cluster shown in Fig.~\ref{fig_exch}. We have used in addition the reflection 
symmetries of the cluster for $k = 0$ and $k = \pi$. For the description of 
the INS results, and to test whether the spin of the ground state is $S = 10$, 
one must consider at least the sectors with $S_z \ge 9$. The largest matrix 
dimension is then 324~908, which occurs for $S_z = 9$, $k = \pi/2$. This 
dimension is sufficiently small that rapid diagonalization by the Lanczos 
procedure is possible on a modern personal computer. Indeed, fast 
diagonalization is a necessary condition for an analysis requiring the 
consideration of many sets of parameters $\{J_i\}$. Finally, using our 
optimal parameter set we have performed further computations with $S_z = 
0$ to ensure that the spin of the ground state is indeed $S = 10$. For 
$S_z = 0$ the largest dimension to be considered is 1~073~763 when using 
spin inversion symmetry.

The condition of a ground state with spin $S = 10$ already
sets a strong constraint on the ratios of the exchange interactions,
and, for example, may be used to exclude the parameter sets proposed in
Refs.~\onlinecite{Sessoli93a} and \onlinecite{Boukhvalov02}.
We have computed low-energy spectra for approximately 1500 independent
parameter sets which satisfy the $S = 10$ ground-state condition. Many of 
these sets were considered in Ref.~\onlinecite{Regnault02}, but we have 
performed independent computations for all parameter sets and added 
further points in regions indicated by the susceptibility calculation 
as potentially relevant. We note that the $S = 10$ condition is invariant 
under rescaling of the interaction parameters, whence one obtains a line 
in the four-dimensional parameter space from each independent 
diagonalization.

\subsection{Magnetic Susceptibility}

\label{secSusc}

The behavior of the static susceptibility of a magnetic material at high 
temperatures may be used in a systematic manner to extract information 
concerning the exchange interactions within the system (see for example 
Ref.\ \onlinecite{Schmidt01}). The simplest example of this process is 
that the first correction to paramagnetic behavior in a FM or AFM system 
gives the Curie-Weiss temperature, which is directly proportional to the 
sum of the couplings to each individual spin. In Mn$_{12}$-acetate the 
application of this analytical approach is complicated by two factors: 
first, the susceptibility is still far from its high-temperature limit 
when the sample decomposes above 300 K, and second, a number of the 
coupling coefficients to be determined may be of similar magnitude.

\begin{figure}[t!]
\centerline{\includegraphics[width=8 true cm]{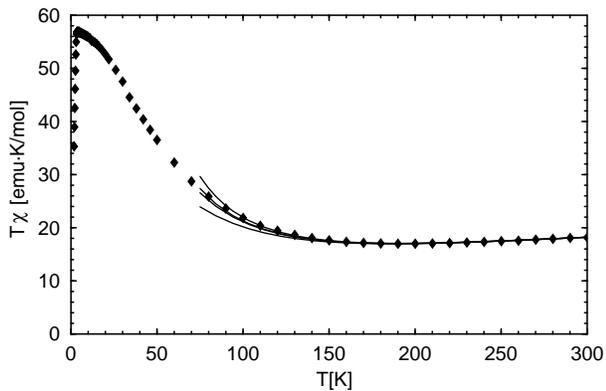}}
\caption{Static magnetic susceptibility measured with an applied field of 
0.1 T (points) and the best fit by obtained by high-temperature series 
expansion (solid lines). The parameter set used is given in eq.~(\ref{eqJfit}).
The solid lines are obtained from different approximations to the 
8th-order high-temperature series, and represent the error estimate 
for the extrapolation procedure.}
\label{fig:susc}
\end{figure}

We begin by showing our measurements of the magnetic susceptibility of 
Mn$_{12}$-acetate, which are given by the symbols in Fig.~\ref{fig:susc}. 
This measurement was performed on a 50 mg powder of the fully deuterated 
sample, over a temperature range of $1.8$-$300$ K and with an applied 
field of 0.1~T, using a Quantum MPMS XL-5 magnetometer, and the data were
corrected for diamagnetic contributions. Our data is very similar to 
that obtained in Ref.~\onlinecite{Schake94}, with the exception of the 
dip at the lowest temperatures, which is significantly sharper in our 
0.1~T measurement because the spread of sublevels in the $S = 10$ ground 
state is narrower than in an applied field of 1~T.\cite{Schake94} We 
comment that the presence of a small component of a faster-relaxing 
species of Mn$_{12}$-acetate molecules, estimated at 3-4\% in 
Ref.~\onlinecite{Mirebeau99}, would be visible only at the lowest 
temperatures.

The measured susceptibility is then fitted in the form 
\begin{equation}
T \chi(T) = \sum_{n=0}^\infty \tilde{C}_n\, T^{-n} \, .
\label{ehte1}
\end{equation}
Making use of the $g$-values obtained for Mn$_{12}$-acetate by EPR 
measurements on a powder sample,\cite{Barra97} we begin by fixing the 
value of the leading coefficient as $\tilde{C}_0 = 30.5$ emu K mol$^{-1}$. 
The higher coefficients $\tilde{C}_n$, $n \ge 1$, are then determined from 
a fit to the high-temperature part of the data, from which one may obtain 
an estimate of the uncertainty in the coefficients from their variation
with the temperature range used for the fit.

The interaction parameters $J_1, J_2, J_3$, and $J_4$ are then fitted
by comparing the coefficients $C_n$ and $\tilde{C}_n$ in Eqs.~(\ref{ehte2}) 
and (\ref{ehte1}). Because this is an overconstrained problem, and in order 
to satisfy the condition that the ground state be $S = 10$, we have 
optimized a suitability function. This function was in turn defined as
a sum of differences of appropriately rescaled coefficients $C_n$ and 
$\tilde{C}_n$, $1 \le n \le 6$, normalized by the estimated uncertainties 
in the latter coefficients. The single most important piece of information 
which can then be extracted is $\tilde{C}_1$, which is given by a weighted 
sum of the interactions (see Eq.\ (\ref{eqCval})) and hence sets the overall 
energy scale of the problem.

\begin{figure}[t!]
\centerline{\includegraphics[width=7 true cm]{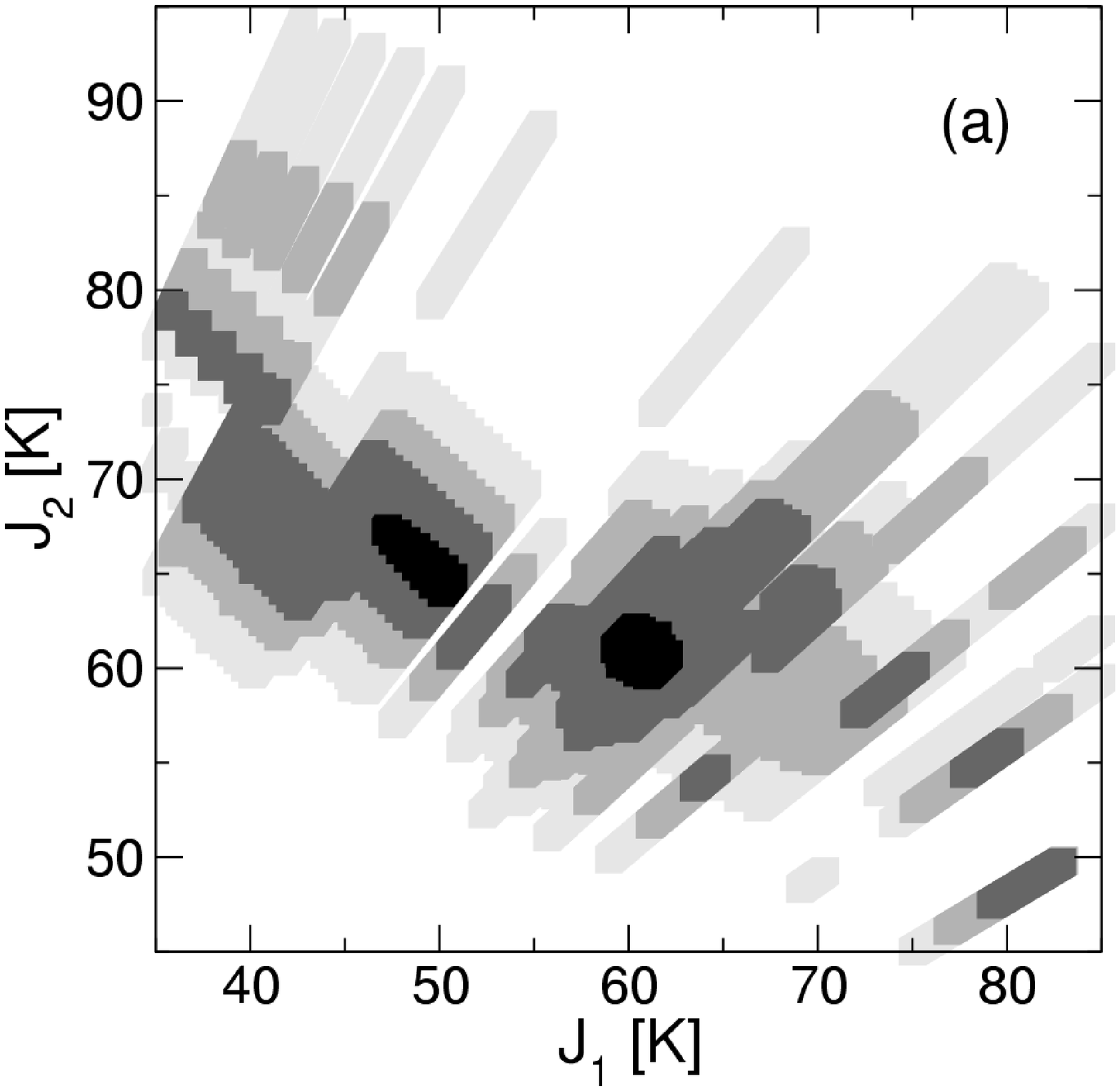}} 
\medskip
\centerline{\includegraphics[width=7 true cm]{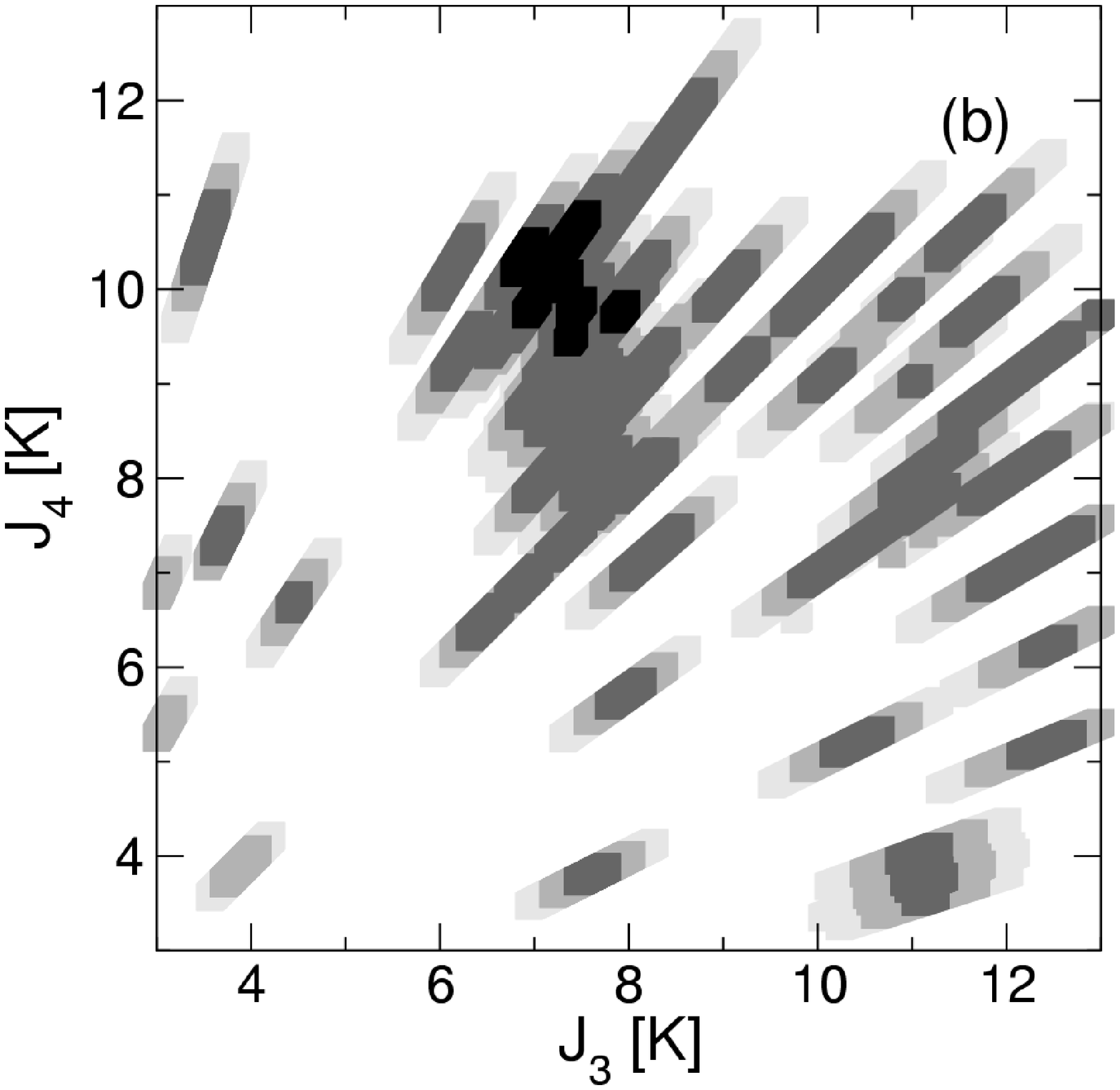}}
\caption{Suitability of parameter sets giving a ground state of spin 
$S$ = 10 for fitting the high-temperature susceptibility data: (a) as 
a function of $J_1$ and $J_2$ for all values of $J_3$ and $J_4$; (b) 
as a function of $J_3$ and $J_4$ for all values of $J_1$ and $J_2$. 
Darker shades indicate superior agreement with $\chi(T)$.}
\label{fig:gschifit}
\end{figure}

The result of this analysis for fixed $\tilde{C}_0 = 30.5$ emu K 
mol$^{-1}$ is shown in Fig.~\ref{fig:gschifit}. Each point in this 
figure is a projection of a point in the four-dimensional parameter 
space yielding an $S = 10$ ground state. Fig.~\ref{fig:gschifit}(a)
shows a projection of the four-dimensional parameter space onto the 
$(J_1,J_2)$ plane and Fig.~\ref{fig:gschifit}(b) onto the $(J_3,J_4)$ 
plane. The suitability of the parameter sets for describing the measured 
susceptibility $\chi(T)$ is shown by shading of the points, where darker 
shades denote higher and lighter shades lower levels of agreement. In 
Fig.~\ref{fig:gschifit}(a) we observe two maxima of suitability. The left 
maximum gives rise to $S = 11$ excitations with energies of approximately 
250~K ($\approx 21.5$ meV), an energy range in which from Sec.~\ref{secAss} 
no such excitations are expected, and thus we focus on the other maximum.
We conclude that the requirements for an $S = 10$ ground state, for matching
the susceptibility $\chi(T)$, and for $S = 11$ excitations of sufficiently 
high energies, constrain the exchange parameters (to within uncertainties of 
order 3--4 K), to the values $J_1 \approx J_2 \approx 61$ K, $J_3 \approx 
7.5$ K, and $J_4 \approx 10$ K (Fig.~\ref{fig:gschifit}). 

In this region of parameter space, we have further optimized a definitive 
set of exchange interactions by considering the magnetic excitations, 
which are discussed in detail in Sec.~\ref{secAspec}. The primary constraint
set by the INS measurements presented in Sec.~\ref{secExp} is the requirement
that the lowest $S = 11$ excitation have an energy of at least 285~K
($\approx 25$ meV). This condition moves the optimal parameter set away from
the minima of the suitability function found in Fig.~\ref{fig:gschifit} for 
the illustrative fitting procedure with fixed $\tilde{C_0}$, yielding the set 
\begin{eqnarray}
J_1 = 67.2~{\rm K}\ (5.79~{\rm meV}), &&\!\!\!\!
J_2 = 61.8~{\rm K}\ (5.33~{\rm meV}), \nonumber \\
J_3 = 7.8~{\rm K}\ (0.67~{\rm meV}), &&\!\!\!\!
J_4 = 5.6~{\rm K}\ (0.48~{\rm meV}). \label{eqJfit}
\end{eqnarray}

We will focus henceforth on this parameter set, and begin by justifying it 
as our final choice. For this purpose we present a direct comparison of the 
magnetic susceptibility obtained with the parameters of Eq.~(\ref{eqJfit}) 
and the experimental data. Instead of fixing the precise value of the
zeroth-order coefficient we use this as a further fitting parameter, 
obtaining the slightly modified coefficient $\tilde{C}_0 = 29.5$ emu K 
mol$^{-1}$. This value corresponds to an effective average $g$-factor 
$g_{\rm eff} = 1.935$, and the modification can be interpreted as arising 
from single-ion anisotropies which are not otherwise present in the 
analysis (Sec.~\ref{secAss}). In addition, we now use all eight orders 
of the high-temperature expansion, two more than were used in the 
calculation of the suitability function shown in Fig.~\ref{fig:gschifit}. 
The lines in Fig.\ \ref{fig:susc} show several Pad\'e approximants to the 
series obtained from the optimal parameter set. Without entering into the
technical details of this procedure, we note only that the different
Pad\'e approximants to the series allow an estimate of the uncertainty
in the extrapolation of the high-temperature expansion. The divergence
of the different approximants below 150 K, and the departure from the 
data at low temperatures, are to be expected in a high-temperature series 
expansion. Within the uncertainty of the extrapolation, we find good
agreement with the data over the entire temperature range from 
approximately 80~K to 300~K. In the high-temperature regime which is 
relevant for our analysis, the effect of a small concentration of 
faster-relaxing species, i.e.~of molecules with slightly different 
exchange constants, would be negligible both in the suitability of the 
fit and in comparison with the other sources of uncertainty. We stress 
that the overall agreement is in fact better than that provided by the 
maxima in Fig.~\ref{fig:gschifit} under the constraint $\tilde{C}_0 = 
30.5$ emu K mol$^{-1}$. We conclude that the optimized parameters of 
Eq.~(\ref{eqJfit}) provide a good description of the experimental data 
for the magnetic susceptibility.

\subsection{High-field Magnetization}

\label{secHF}

\begin{figure}[t!]
\centerline{\includegraphics[angle=0,width=7 true cm]{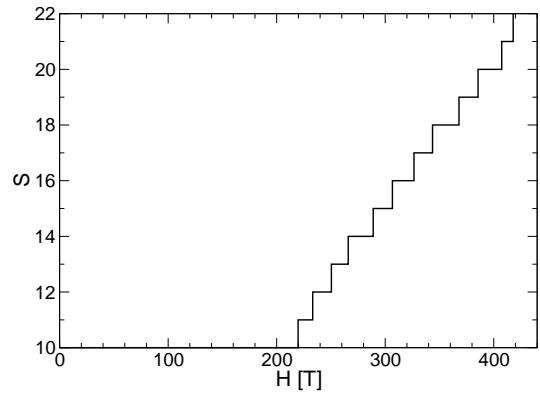}}
\caption{Calculated magnetization curve for a Mn$_{12}$-acetate cluster 
with exchange parameters given in eq.~(\ref{eqJfit}).}
\label{fig:magn}
\end{figure}

We comment here that the constraint on the overall magnitude of the 
interaction parameters provided by fitting the susceptibility requires
a new interpretation of the high-field magnetization 
data.\cite{Zvezdin98,Gatteschi_review02} The fixed value of the weighted 
sum of exchange constants $C_1$ has a direct correspondence to the 
saturation field of the system, at which all bonds must be polarized 
ferromagnetically. The predicted $T = 0$ magnetization curve, by which 
is meant here the spin $S \sim M / g \mu_{\rm B}$ as a function of 
applied field $H$, is shown in Fig.~\ref{fig:magn} for the parameters 
of Eq.~(\ref{eqJfit}), using the same value $g_{\rm eff} = 1.935$ as for 
the susceptibility to express the magnetic field $H$ in units of T. The 
low-field data is consistent with that of 
Refs.~\onlinecite{Zvezdin98,Gatteschi_review02},
in that the $S$ = 10 ground state persists up to fields in excess of
200 T. The relatively large number of narrow steps in the calculated
magnetization may then also be compared with the 
regularly spaced steps in the data. The location of the first step, 
corresponding to the transition between $S = 10$ and $S = 11$ ground 
states, may not be underestimated from our computation by more than 
10--20 T. The optimal susceptibility fit therefore implies peaks in the
$dM/dH$ data at fields smaller than the lowest peaks assigned and used to
determine the exchange parameters $\{J_i\}$ in Ref.\ \onlinecite{Regnault02}.
However, we believe that the raw experimental
data\cite{Zvezdin98,Gatteschi_review02} do not 
in fact exclude further peaks in the region between 200 and 300 T.

The uncertainty due to errors in the exchange constants may accumulate
to a value on the order of 100 T for the saturation field, which is the 
transition between $S=21$ and $22$ ground states, but larger deviations 
become increasingly unlikely given the constraint on the overall energy 
scale obtained from the susceptibility. Hence the single largest peak in 
the $dM/dH$ data of Refs.~\onlinecite{Zvezdin98,Gatteschi_review02},
which appears at 520~T, 
occurs at or above the predicted saturation. Indeed, none of the 
magnetization curves which we have computed yield indications for 
a single step which would be substantially more pronounced than all of 
the other steps. Given the reliability of a susceptibility measurement, 
and the destructive nature of the explosive compression technique involved
in the high-field magnetization measurement, we suggest that the former 
is more representative. In this interpretation, all signals at and beyond 
the largest peak at 520~T in the $dM/dH$ data would correspond to times 
when the sample or other components of the experimental apparatus are 
already disintegrating.

\subsection{Excitation Spectrum}

\label{secAspec}

\begin{table}[b!]
\caption{Magnetic energy level spectrum obtained for the first 17
nondegenerate states with $S \ge 9$ of a Mn$_{12}$-acetate exchange 
model with the parameters of Eq.~(\ref{eqJfit}), classified according 
to spin state, level degeneracy, and spatial symmetry sector. No 
energetic correction is applied for uniaxial anisotropy of the cluster. }
\vspace{0.5cm} 
\begin{tabular}{|c|c|c|c|}
\hline $\;$ Spin $S$ $\;$ & $\;$ Energy [K] $\;$ & $\;$ Degeneracy & $\;$ 
Symmetry $k$ \\ \hline
10  &  0.00 & 1 & $0\psym{+}$ \\
9   &  28.48 & 2 & $\pm \pi/2$ \\
9   &  44.47 & 1 & $\pi\psym{+}$ \\
9   &  91.46 & 1 & $0\psym{+}$ \\
9   &  119.67 & 2 & $\pm \pi/2$ \\
9   &  159.61 & 1 & $\pi\psym{-}$ \\
11  &  285.58 & 1 & $0\psym{+}$ \\
10  &  293.74 & 2 & $\pm \pi/2$ \\
10  &  297.30 & 1 & $\pi\psym{-}$ \\
11  &  303.23 & 2 & $\pm \pi/2$ \\
9   &  304.45 & 1 & $0\psym{+}$ \\
9   &  304.65 & 1 & $\pi\psym{+}$ \\
9   &  306.73 & 2 & $\pm \pi/2$ \\
9   &  307.14 & 1 & $\pi\psym{-}$ \\
10  &  311.11 & 1 & $0\psym{+}$ \\
10  &  317.28 & 1 & $0\psym{-}$ \\
9   &  324.51 & 1 & $0\psym{+}$ \\
\hline
\end{tabular}
\label{tabSpec}
\end{table}

We turn next to an analysis of the excited states. Table \ref{tabSpec} 
shows the lowest nondegenerate magnetic energy levels for a cluster with 
the optimized set of exchange interactions (\ref{eqJfit}). These levels are 
labeled by their spin and spatial symmetry sectors. Qualitatively, the 
parameter values can be seen to provide a number of low-lying $S$ = 9 
states, as required for comparison with INS data [transitions (I) and (II)], 
without permitting the existence of an $S = 11$ state or additional $S = 10$ 
states below a significantly higher energy, whose value is in agreement 
with the limits set by INS [transition (III)] and high-field 
magnetization measurements. The fact that five nondegenerate $S$ = 9 
states are found in the low-energy manifold may be justified by 
straightforward considerations based on a spin-wave description 
(Sec.~\ref{secSW}), and can be used as an aid to experimental interpretation 
(Sec.~\ref{secAss}). By considering the symmetry of the $S$ = 9 states one 
may also observe which of these would be sensitive to a breaking of 
fourfold cluster symmetry, and thus by comparison with the width of 
the observed INS peaks establish an approximate upper bound on the 
extent of any such departure from symmetry (Sec.~\ref{secSW}). 

\begin{figure}[t!]
\centerline{\includegraphics[width=8 true cm]{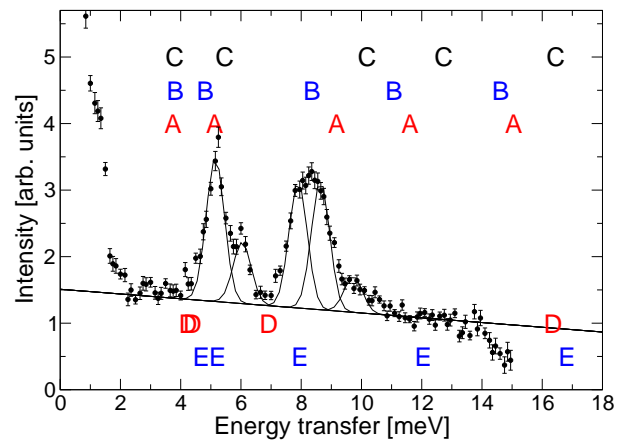}}
\caption{Comparison with INS data of the location of calculated magnetic 
excitations, obtained using the optimal parameter set (\ref{eqJfit}).
An anisotropy correction is applied by using Eq.~(\ref{eaas}) with 
a single $D$ value for all (ground and excited) states of each total 
spin $S$ = 9, 10, or 11. The experimental data was obtained on MARI at 
(a) low energy transfer [Figs.~\ref{fig_mari}(b) and (c)] and (b) high 
energy transfer [Fig.~\ref{fig_mari}(a)].}
\label{fig:ncomp}
\end{figure}

A quantitative comparison between the numerical results and the positions 
of the excited levels observed by INS requires that the single-ion 
anisotropy terms be taken into account. As stated above, we have included 
these terms only at the level of an effective uniaxial anisotropy acting 
on the cluster spin [Eq.~(\ref{eaas})]. The results of this exercise are 
shown in Fig.~\ref{fig:ncomp}. With regard to position, type, and number 
of groups of peaks, the properties of the model are in good accord with 
the measured data sets. Levels (Ia), (Ib) and (IIa) deviate by no more 
than 20~K ($\approx 2$~meV) from the theoretical results while the
discrepancy for levels (IIb) and (IIc) is somewhat larger. In this regard, 
we note that the anisotropy shift is on the order of 15 K ($\approx 1.3$ 
meV) and that rather small changes (of order 1 K) in the values of any 
individual coupling parameter $J_i$ may cause large (of order 10 K), 
albeit not entirely uncorrelated, changes in the positions of the different 
excited states. Further optimization would require a full analysis of the 
uniaxial anisotropy at the single-ion level. Although computation of 
spectra for a number of selected parameter sets with the inclusion of 
single-ion terms is possible with current computing technology, it is a 
demanding numerical task due to the fact that $S_z$ is not a conserved 
quantum number in this case. As a consequence the full optimization of 
parameters would no longer be possible, and for this reason we do not 
pursue single-ion terms further here; the energy scale of the resulting 
mixing and splitting of levels in different manifolds remains small on 
the scale of the variability present in the exchange interactions.
We thus focus on the robust features which may be extracted at the 
current level of refinement, namely a reproduction of the properties of 
the peak groups (I), (II), and (III) (Fig.~\ref{fig:ncomp}).

\begin{table}[b!]
\caption{Incoherent sum of squared matrix elements $M_n = \sum_i \abs{\langle 
g(10,9) | S_i^z|e_n(9,9) \rangle}^2$ between ground and excited states 
for the 7 low-lying $S$ = 9 levels, obtained with the parameters of 
Eq.~(\ref{eqJfit}).}
\vspace{0.5cm} 
\begin{tabular}{|c|c|c|} \hline 
$n$ & $\;$ Energy [K] $\;$ & $M_n$ \\ \hline
$\;\; 1,2 \;\;$ &  28.48 & $\;$ 0.25506 $\;$  \\
$3$ &    44.47 &  0.21426  \\
$4$ &    91.46 &  0.19241  \\
$5,6$ & 119.67 &  0.19485  \\
$7$ &   159.61 &  0.17911  \\ \hline
\end{tabular}
\label{tabInc}
\end{table}

In addition to the position of the magnetic excitations it is important 
also to compute their intensities to ensure that transitions to all of 
the calculated levels are allowed. We have computed the matrix elements 
$\langle g(10,9)|S_i^z|e_n(9,9) \rangle$, which by spin rotation symmetry 
are related directly to the elements of the operator $S_i^-$ appropriate 
for the spin-flipping action on the $S = 10$ ground state $| g(10,10) 
\rangle$ of a neutron-scattering process, for all sites $i = 1, \dots, 
12$ of the cluster. These elements may be used in a calculation including 
the structure factor of the molecule to obtain the transition matrix 
element at any wave vector ${\bm Q}$. In Table \ref{tabInc} we show only 
the incoherent sum (no phase factor) of the squares of the site matrix 
elements: this is sufficient to show that none of the levels is excluded 
for symmetry reasons, and that all may be expected to have similar weights.
However, as a result of the degeneracy of the pairs ($|e_1 \rangle, |e_2 
\rangle$) and ($|e_5 \rangle, |e_6 \rangle$), for INS purposes the relevant 
quantity would be the sum of these two weights. One may then expect two 
peaks whose integrated intensity is double that of the other three, a 
result certainly not inconsistent with the INS data. 

We conclude this section by computing the spin distribution of the 
ground-state wave function. Table \ref{tabGSdens} shows the values of 
$(S_i^z)^2$ on the core $S$ = 3/2 spins ($i$ = 1, 4, 7, 10; Fig.~2), on 
the $S$ = 2 crown spins coupled to one core spin by exchange parameter 
$J_1$ ($i$ = 2, 5, 8, 11), and on the $S$ = 2 crown spins coupled to two 
core spins by exchange parameter $J_2$ ($i$ = 3, 6, 9, 12). The remarkable 
feature of these results is that approximately 90\% of the spin weight 
of the crown spins is in the classical ferrimagnetic state invoked in 
Sec.~\ref{secGS} to justify the $S$ = 10 ground state. This value drops 
only to 80\% for the core spins, whose mutual coupling is weak but 
frustrating. The degree of overlap with the classical state is confirmed 
by considering the $S_z$ correlations between neighboring inequivalent 
sites (Table \ref{tabGSdens}). 

\begin{table}[b!]
\caption{Spin distribution of the ground state of Mn$_{12}$-acetate, 
characterized by the values of $(S_i^z)^2$ on all sites and by spin 
correlations between neighboring inequivalent sites, obtained with the 
parameters of Eq.~(\ref{eqJfit}).}
\vspace{0.5cm} 
\begin{tabular}{|c|c|c|} \hline 
  & 1.77542 & $i$ = 1, 4, 7, 10 \\
$\; \langle g(10,10)|(S_i^z)^2|g(10,10) \rangle \;$ & 3.63387 
& $i$ = 2, 5, 8, 11 \\
  & 3.55449 & $\;$ $i$ = 3, 6, 9, 12 $\;$ \\ \hline
  & $\;$ $-2.40483$ $\;$ & $i$ = 1, $j$ = 2 \\
$\langle g(10,10)|S_i^z S_j^z|g(10,10) \rangle$ & $-2.32373$ & $i$ = 1, 
$j$ = 3 \\
  & 3.45023 & $i$ = 2, $j$ = 3 \\ \hline
\end{tabular}
\label{tabGSdens}
\end{table}

The classical alignment is ensured by two factors. The first is the 
dominance of the exchange couplings $J_1$ and $J_2$, which forces all 
of the core spins be antiparallel to the crown spins, and therefore 
mutually FM. The weaker, frustrating $J_3$ and $J_4$ bonds then cause 
only minor deviations from the classical state. The second is the 
near-equivalence of the interactions $J_1$ and $J_2$, to which the 
sensitivity of the system is shown by the size of the black region in 
Fig.~\ref{fig:gschifit}(a). The effect of a departure from this equivalence 
is illustrated by the example of the $J_1$-dominated parameter set 
originally proposed in Ref.~\onlinecite{Sessoli93a}, where sites $i$ = 1, 
2, 4, 5, 7, 8, 10, and 11 would be coupled to form effective $S$ = 1/2 
units, and only very small ordered spin components would be observed 
on these. Precisely because corrections of this form are small in 
Mn$_{12}$-acetate, one may expect that a spin-wave description of the 
excited states\cite{Yamamoto02} is in fact meaningful despite the small 
size of the system. Finally, the relative lack of quantum mechanical 
fluctuation effects may also be ascribed in part to the ``large'' values 
of the ionic spins ($S$ = 3/2 or 2, as opposed to $S$ = 1/2). 

\subsection{Spin-wave Analysis}

\label{secSW}

\begin{figure}[t!]
\centerline{\includegraphics[width=7.5 true cm, angle = 0]{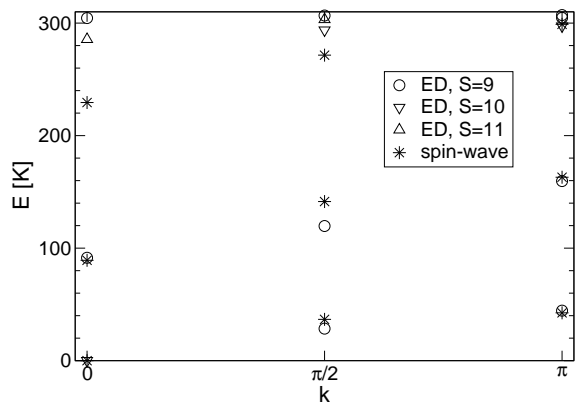}}
\caption{Comparison between magnetic excitations calculated by exact 
diagonalization (ED) and within a spin-wave approximation using the 
parameter set of Eq.~(\ref{eqJfit}). No uniaxial anisotropy is included. 
Higher $S$ = 9 and 10 excitations not included within the 3 spin-wave 
branches enter only at the highest energy of the $S$ = 11 branch. }
\label{fig:sw}
\end{figure}

We have found that the optimal parameter set for Mn$_{12}$-acetate 
yields a ground state whose spin distribution can be said to be 
``80-$90$\% classical''. This result, which may seem surprising for 
a relatively small, quantum system, is a consequence of the fact that 
the two dominant exchange interactions, $J_1$ and $J_2$, are unfrustrated. 
The observation that an $S$ = 10 ground state may be found only in a very 
small region of four-dimensional parameter space around this rather 
special limiting case is evidence that any significant frustrating 
interactions $J_3$ or $J_4$, or deviations from $J_1 \approx J_2$, 
would in fact destroy the classical ground state. The gap to the lowest 
excited states of the exchange model without anisotropy, measured as 
being approximately 4 meV (40K) in Mn$_{12}$-acetate [Eq.~(\ref{eea}), 
Fig.~\ref{fig_diagram}], may also be considered as a sensitive indicator 
of the proximity of any similar parameter set to a ground state of a 
different spin $S \ne 10$. 

One consequence of this quasi-classical nature is that the magnetic 
excitations may be considered by conventional spin-wave theory. Within 
this treatment, first presented in Ref.~\onlinecite{Yamamoto02}, the cluster 
is considered as a ring of 4 unit cells each with 3 inequivalent sites. 
Two of the spin-wave branches correspond to reduction of the spin 
component $S_z$ = 2 on the crown sites, leading to an $S$ = 9 state 
of the classical cluster, while the third corresponds to an increase 
of the component $S_z = - 3/2$ on the core sites, leading to an $S$ 
= 11 state. The 4-cell ring has 3 inequivalent $k$ points, where $k$ 
is an effective wave vector around the ring, whence one expects 9 
levels in 3 branches: the ground state, 5 $S$ = 9 states, and 3 $S$ = 
11 states. The exact energy levels for any parameter set $\{J_i\}$ may 
be computed by exact diagonalization, and in Fig.~\ref{fig:sw} we show the 
comparison between these levels and the spin-wave approximation for the 
optimal Mn$_{12}$-acetate set (\ref{eqJfit}). The agreement is remarkably 
good for such a crude approximation to a small quantum system, again 
because of the special, ``classical'' set of exchange interactions 
found in Mn$_{12}$-acetate. The authors of Ref.~\onlinecite{Yamamoto02} 
applied their approximate treatment to the parameter set of Sessoli 
{\it et al.},\cite{Sessoli93a} for which the ground state deviates 
strongly from the classical paradigm, and unsurprisingly a correspondence 
between their excited states and the exact results is difficult to 
establish in this case. Returning to the interpretation of the INS data, 
the peak groups I and II (see Table \ref{tabSum}) can be identified 
respectively with the first and second branches of spin-wave excitations, 
all of which have spin $S = 9$ and provide the appropriate symmetry sectors 
($k$ in Table \ref{tabSpec}). 

The success of the spin-wave approximation in the physical parameter 
regime may be used for two further purposes. One is to investigate 
the sensitivity of the computed excitations to changes in individual 
interaction parameters without the need for a lengthy numerical 
calculation. We have performed this exercise in order to obtain the 
optimized parameter set (\ref{eqJfit}), and also to observe the relative 
changes of different excited levels with the parameters. Here the fact 
that certain excitation energies change together suggests that a full, 
independent tuning of the levels to obtain exact agreement with the 
measured INS peaks may not be possible within a four-parameter exchange 
model, and that this would probably require the reintroduction of 
microscopic anisotropy terms neglected in the current analysis. In 
this connection we note only that single-ion anisotropy terms are 
expected to make the spin-wave branches less dispersive, which would 
improve the agreement with the INS results. The second purpose is to 
investigate the consequences of a breaking of fourfold cluster symmetry, 
if in fact one were present. In this case $J_{2a} \ne J_{2b}$ and $J_{4a} 
\ne J_{4b}$, and the degeneracy of the $k = \pm \pi/2$ states in Table 
\ref{tabSpec} would be lifted. With reference to Table \ref{tabInc}, 
these are the degenerate level pairs $(|e_1 \rangle, |e_2 \rangle)$ and 
$(|e_5 \rangle, |e_6 \rangle)$. From the fact that the INS data may 
already be fitted by 5 Gaussian peaks of rather similar widths 
[$\Gamma = 0.77$ meV (8.9 K) in (I) and (II)] not far from the 
resolution limit, the spin-wave description can be used to set upper 
bounds on $\Delta J_2 = |J_{2a} - J_{2b}|$ and $\Delta J_4 = |J_{4a} 
- J_{4b}|$ of approximately 0.1 meV (1 K).

\section{Summary}

\label{secSum}

We have performed inelastic neutron scattering measurements up to high 
energies to identify and characterize the magnetic excitations of 
Mn$_{12}$-acetate. We find that all of the lowest energy levels, which occur 
in two groups at 5--6.5 meV (60--75 K) and 8--10.5 meV (95--120 K), appear 
to be $S = 9$ states. There are approximately five such levels, and their 
intensities do not vary by more than a factor of $2$--$3$. Higher levels, 
including the lowest $S = 11$ states, are found at a significantly higher 
energy, namely the newly identified transition peak (III) at 27 meV (310 K), 
which is qualitatively consistent with the location of the first 
magnetization step. Taken together with the fact that the ground state has 
spin $S = 10$, the qualitative features of the measurements are already 
sufficient to restrict the interaction parameters of the cluster almost to 
a unique set. All of the available experimental information is reproduced 
by a parameter set with $J_1 \sim J_2 \sim 5.5$ meV (65 K), while $J_3$ 
and $J_4$ are smaller than 1 meV (10 K); refinement of the more robust 
quantitative features leads to the parameter set given in Eq.~(\ref{eqJfit}).

Essential supplementary fitting information is provided by the magnetic 
susceptibility, which we have computed by a systematic high-temperature 
series expansion. A correct reproduction of the high-temperature limit 
sets a constraint on the four interaction parameters which determines 
the overall energy scale of the couplings involved. In this context we 
note that the leading term may be expected from the cluster geometry 
and dominant couplings to be determined by the combination $J_1 + 2 
J_2$, which is indeed very close to the value $J_1$ = 216 K extracted 
in Ref.~\onlinecite{Sessoli93a} on the assumption of one dominant 
interaction. In fact this constraint is difficult to reconcile with 
the high-field magnetization data, whose saturation field should be 
given by a similar combination of terms. On the basis of our 
measurements and calculations, the largest feature in the $dM/dH$ curve 
of Refs.~\onlinecite{Zvezdin98} and \onlinecite{Gatteschi_review02} 
should correspond qualitatively to the saturation field. 

In the consistent parameter set all four of the exchange interactions 
are AF. From qualitative considerations based on exchange and superexchange 
processes these results are fully plausible, despite the proximity of 
certain $\mu$-oxo bridging angles to 90$^\circ$. The quantitative similarity 
of $J_1$ and $J_2$ in spite of the very different types of exchange pathway 
involved, as well as the small but positive value of $J_3$, emphasize 
the difficulties inherent in performing {\it ab initio} calculations 
sufficiently accurate to reproduce the physical properties of complex 
magnetic systems. Even small deviations from these values are sufficient 
to change the spin of the ground state (Sec.~\ref{secSusc}), to bring 
excluded higher-spin states into the low-energy manifold, or to return 
a high-temperature susceptibility well outside the limits imposed by 
the experimental measurement. 

Throughout our analysis we have focused on the qualitative features which 
a candidate parameter set must reproduce. At the quantitative level there 
are two sources of error: the first is the uncertainty in the multiparameter 
problem of fitting the set $\{J_i\}$ to disparate (and occasionally 
conflicting) pieces of data which are themselves subject to measurement 
errors; the second is the terms in the Hamiltonian which we have neglected, 
primarily exchange asymmetries, single-ion anisotropies and higher-spin 
interactions. Because of the first, it is manifestly not meaningful to 
ascribe parameter values to the second, and yet these would be required 
for a perfect quantitative account of an ideal magnetic excitation 
spectrum. Thus we have restricted our considerations to robust results 
and qualitative fitting, adequate at the level of groups of INS peaks. 
These conditions are in fact sufficient to identify an unambiguous 
set of exchange constants. Although for reasons of simplicity we have 
focused our presention on the precise values given in Eq.~(\ref{eqJfit}), 
we have performed similar analyses for small modifications of this set.
From the variability of exchange parameters between sets yielding a 
description of similar quality, we estimate an error bar of 2--3 K for 
each individual exchange parameter $J_i$. 

In conclusion, measurements of its magnetic properties allow us to 
establish a definitive set of intramolecular exchange interactions 
for Mn$_{12}$-acetate. The only consistent parameter set is
\begin{eqnarray}
J_1 & = & 67 \pm 3 {\rm K} \;\; (5.8 \pm 0.3 {\rm meV}), \nonumber \\
J_2 & = & 62 \pm 3 {\rm K} \;\; (5.3 \pm 0.3 {\rm meV}), \nonumber \\
J_3 & = & 8 \pm 3 {\rm K} \;\; (0.7 \pm 0.3 {\rm meV}), \nonumber \\
J_4 & = & 6 \pm 3 {\rm K} \;\; (0.5 \pm 0.3 {\rm meV}), \nonumber 
\label{ocps}
\end{eqnarray}
where the error bars denote the order of uncertainty for each $J_i$. 
These parameters provide both a straightforward understanding of the 
ground state and low-energy spectrum, and a resolution of the conflicts 
in the existing literature.

\begin{acknowledgments}

We are grateful to D.T. Adroja (ISIS, UK), H. Mutka (ILL, France), and 
S. Janssen (PSI, Switzerland) for assistance during the INS experiments. 
We thank N.\ Regnault and T.\ Jolicoeur for helpful provision of data 
concerning the spin sector of the ground state in different parameter 
regimes. Numerical calculations of the high-temperature series were performed 
on the machine {\tt cfgauss} at the computing center of the TU Braunschweig. 
This work was supported by the Swiss National Science Foundation, the TMR 
programme Molnanomag of the European Union (No: HPRN-CT-1999-00012), and by 
the Deutsche Forschungsgemeinschaft through grant SU 229/6-1. A.H.~would 
like to acknowledge the hospitality of the Institute for Theoretical Physics 
of the University of Hannover, where part of this work was performed. 

\end{acknowledgments}

\end{document}